\documentclass[aps, prl, reprint, superscriptaddress, longbibliography]{revtex4-2}

\usepackage[utf8]{inputenc}
\usepackage{derivative}
\usepackage{graphicx}
\usepackage{dcolumn}
\usepackage{bm}
\usepackage{siunitx}
\usepackage{mathtools}
\usepackage[colorlinks,allcolors=blue]{hyperref}
\usepackage[capitalise]{cleveref}
\usepackage{braket}
\usepackage{booktabs}
\usepackage{nicefrac} 
\newcommand{\hamil}[1][]{\hat{H}_\mathrm{#1}}
\newcommand{\op}[2][]{\hat{#2}_{#1}}
\DeclareSIUnit\angstrom{\text {Å}}

\begin{document}

\title{Substrate polarization alters the Jahn--Teller effect in a single molecule}
\author{Moritz Frankerl}
\email{moritz.frankerl@dipc.org}
\affiliation{Institute for Theoretical Physics, University of Regensburg, 93040 Regensburg, Germany}
\affiliation{Donostia International Physics Center (DIPC), E-20018 Donostia-San Sebasti\'an, Spain}
\author{Laerte L. Patera}
\affiliation{Institute for Experimental Physics, University of Regensburg, 93040 Regensburg, Germany}
\affiliation{Department of Physical Chemistry, University of Innsbruck, 6020 Innsbruck, Austria}
\author{Felix Giselbrecht}
\affiliation{Institute for Experimental Physics, University of Regensburg, 93040 Regensburg, Germany}
\author{Thomas Frederiksen}
\affiliation{Donostia International Physics Center (DIPC), E-20018 Donostia-San Sebasti\'an, Spain}
\affiliation{IKERBASQUE, Basque Foundation for Science, E-48013, Bilbao, Spain}
\author{Jascha Repp}
\email{jascha.repp@ur.de}
\affiliation{Institute for Experimental Physics, University of Regensburg, 93040 Regensburg, Germany}
\author{Andrea Donarini}
\email{andrea.donarini@ur.de}
\affiliation{Institute for Theoretical Physics, University of Regensburg, 93040 Regensburg, Germany}
\date{\today}

\begin{abstract}
    Charge-state transitions of a single Cu-phthalocyanine molecule adsorbed on an insulating layer of NaCl on Cu(111) are probed by means of alternate charging scanning tunneling microscopy. Real-space imaging of the electronic transitions reveals the Jahn--Teller distortion occurring upon formation of the first and second anionic charge states. The experimental findings are rationalized by a theoretical many-body model which highlights the crucial role played by the substrate. The latter enhances and stabilizes the intrinsic Jahn--Teller distortion of the negatively charged molecule hosting a degenerate pair of single-particle frontier orbitals. Consequently, two excess electrons are found to occupy, in the ground state, the \emph{same} localized orbital, despite a larger Coulomb repulsion than the one for the competing delocalized electronic configuration. Control over the charging sequence by varying the applied bias voltage is also predicted.        
\end{abstract}

\maketitle
Scanning tunneling microscopy (STM) is an excellent tool for investigating single molecules on surfaces, as it combines atomic-scale imaging with spectroscopy. 
The hindrance of hybridization with the underlying metal achieved by ultrathin polar insulating films of 2-3 monolayers (ML) has been used to access near pristine electronic structures of molecules \cite{Repp2005},  many-body correlations \cite{Schulz2015, Yu2017}, charge-state lifetimes \cite{Kaiser2023}, coherent spin control \cite{Willke2021} and bistability between $\pi$-diradical open- and closed-shell states \cite{Mishra2024}. Furthermore it enabled the demonstration of fluorescence \cite{Qiu2003, Zhang2017a}, vibronic spectroscopy \cite{Doppagne2017}, electrofluorochromism \cite{Doppagne2018}, switching of the excitonic state \cite{Dolezal2020} in single-molecules as well as energy transfer in molecular dimers \cite{Imada2016, Cao2021}. 
With few exceptions, investigating multiple charge states requires 
inhibiting electron exchange with the underlying substrate and, hence, working on thicker insulating films ($>$20 ML), for which conventional STM is not applicable. 
Steering electron transfer with the tip and probing this transfer with atomic force microscopy (AFM) provides access to multiple charge states \cite{Bussmann2006, Gross2009, Steurer2015} and enabled single-electron transfer between molecules \cite{Steurer2015a, Scheuerer2020}, reorganization energy upon charging \cite{Fatayer2018}, excited molecular states and their lifetimes \cite{Fatayer2021, Peng2021}
to be studied.\\
\indent Alternate charging scanning tunneling microscopy (AC-STM) makes use of this development. In AC-STM single electrons are periodically tunneled back and forth between tip and 
molecule and detected {\it via} AFM.
The signal represents the tunneling probability and can be used to spatially map different charge state transitions
\cite{Patera2019,Patera2019a,Patera2020}.
\begin{figure}
	\centering
	\includegraphics[width=0.85\linewidth]{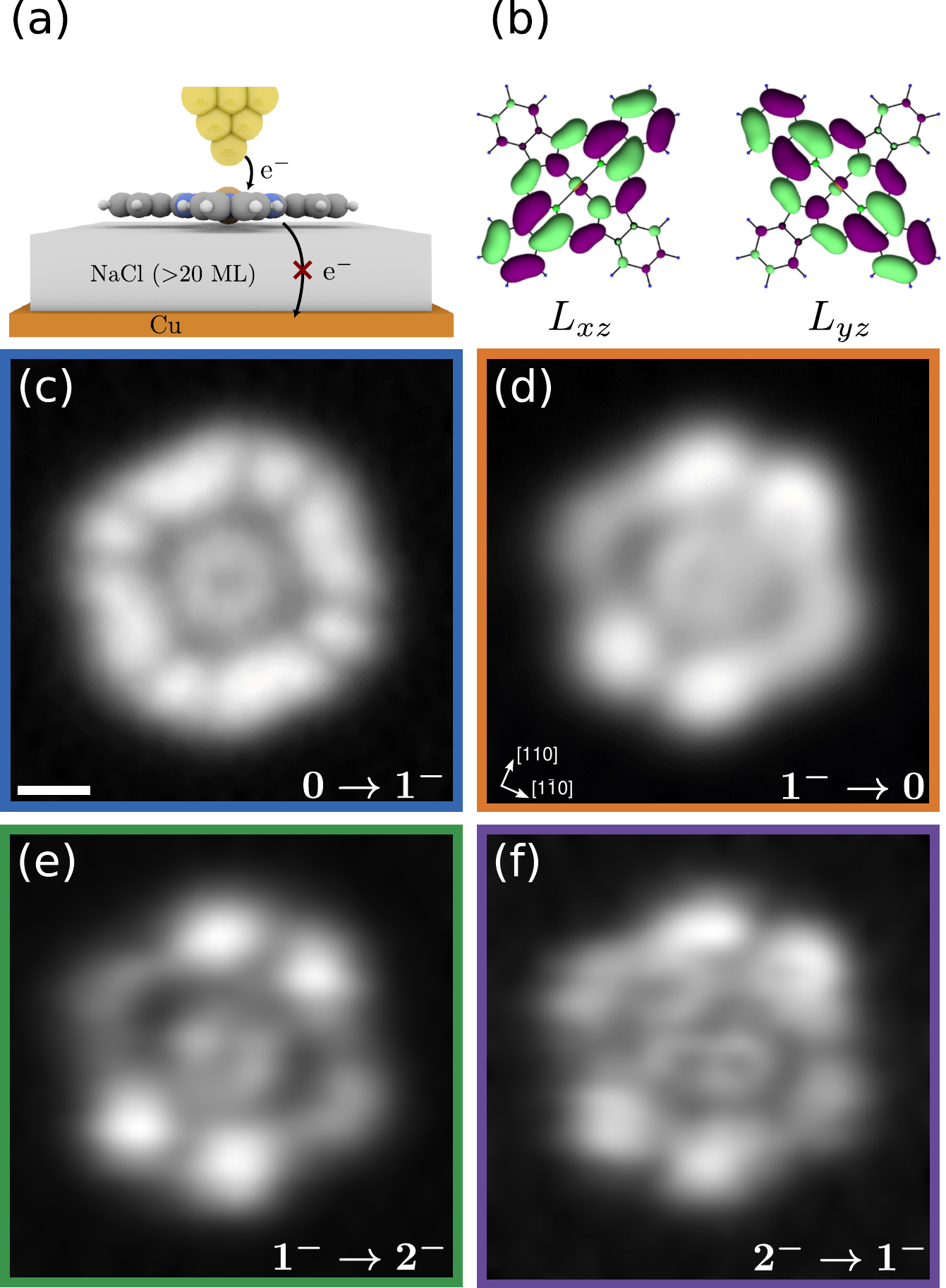}
	\caption{(a) Sketch of the experimental setup. On thick NaCl films ($>$20 ML), tunneling from the CuPc molecule to the Cu substrate is inhibited, enabling charge-state manipulation and the visualization of electronic transitions via AC-STM. 
		(b) Isosurfaces of the two degenerate LUMOs present in CuPc.
	(c-f) AC-STM images (oscillation amplitude $A=1$ Å) corresponding to:
	(c) $0\rightarrow 1^-$ transition ($V_{\rm dc} = \SI{1.0}{\V}$, $V_{\rm ac} = 0.75$ V peak-to-peak (V$_{\rm pp}$), $\Delta z = 3.7$ \AA )
    (d) $1^-\rightarrow 0$ transition ($V_{\rm dc} = \SI{1.0}{\V}$, $V_{\rm ac} = \SI{1.0}{\V}_{\rm pp}$, $\Delta z = 4.4$ \AA)
	(e) $1^-\rightarrow 2^-$ transition ($V_{\rm dc} = \SI{2.3}{\V}$,  $V_{\rm ac} = \SI{0.9}{\V}_{\rm pp}$, $\Delta z = 4.8$ \AA)
	(f) $2^-\rightarrow 1^-$ transition ($V_{\rm dc} = \SI{2.3}{\V} $, $V_{\rm ac} = \SI{1.0}{\V}_{\rm pp}$, $\Delta z = 4.0$ \AA)
         $\Delta z$ are given with respect to an AFM set point of $\Delta f = -1.5$ Hz at $V_{\rm dc} = 0$ V. Scale bar: 5 \AA . 
         (a, c, and d) Reprinted figures with permission from Ref. \cite{Patera2019a}. Copyright 2019 by the American Physical Society.
}
	\label{fig:exp}
\end{figure} 
The use of polar insulators in scanning probe microscopy leads to a strong stabilization of excess charges in adsorbed atoms and molecules \cite{Repp2004, Wu2008, Gross2009, Uhlmann2013, Steurer2015, Patera2020, HernangomezPerez2020}. It also facilitates orbital localization through self-trapping of polarons \cite{Bin2006, Svetin2014}, thus revealing that the insulator's influence reaches beyond mere charge stabilization.\\
\indent Here we show how substrate polarization alters the Jahn--Teller (JT) effect in a charged Cu-phthalocyanine (CuPc) molecule.
While the JT effect is well studied for 
singly-charged CuPc, 
AC-STM reveals that charging with a second electron does not restore the original degeneracy of orbitals but instead leads to a double occupancy of one of the formerly degenerate orbitals.
Our theoretical analysis reveals that the polarizable NaCl substrate does not only strongly enhance the JT splitting, but leads to a qualitatively 
different behavior for the di-anionic species by tipping the energy balance.
Moreover, we predict the transition rate to be dominated by the delocalized configuration at higher bias due to state multiplicity.\\
\indent \emph{Experiments}. We employ AC-STM [\cref{fig:exp}(a)], cf. Supplemental Material (SM) for details~\cite{SM} (see also Ref.~\cite{Giessibl2000} therein), to image the spatial distribution of charge-state transitions for the neutral, anionic and dianionic CuPc species.
To the substrate a DC voltage $V_\text{dc}$ is applied, corresponding to the 
degeneracy point of the desired charge-state transition. Superimposed 
voltage pulses $V_\text{ac}$ drive the charge-state transitions \cite{Patera2019,Patera2019a,Patera2020}.
The map of the 0 $\rightarrow$ 1$^-$ transition [\cref{fig:exp}(c)] exhibits a 4-fold rotational symmetry, similar to the anionic resonance observed by STM on ultrathin NaCl layers (2 ML)  \cite{Uhlmann2013}.
Additional back and forth toggling of the molecule's azimuthal orientation yields, on average, the superposition of two images rotated of about 32 degrees with respect to each other \cite{miwa2016effects, Patera2019a}).
As reported previously~\cite{Patera2019a}, for the reverse transition (1$^-$ $\rightarrow$ 0)[\cref{fig:exp}(d)] the symmetry reduces from 4- to 2-fold due to the JT distortion, which favors in the anion the occupation of one single LUMO \cite{Uhlmann2013}.
Expanding on previous studies we investigate here also the dianionic CuPc.

In the dianionic ground state the second excess electron could either occupy the other (empty) LUMO or further populate the already occupied one.
The JT distortion lowers the energy of the already occupied orbital and thus pushes  the system towards having two electrons in the already occupied LUMO.
However, as can already be estimated from the LUMO densities in \cref{fig:exp}(b) the Coulomb repulsion will be higher for two electrons in one LUMO with respect to electrons being distributed between the LUMOs.
Comparing the AC-STM images of the 1$^-$ $\rightarrow$ 2$^-$ and 2$^-$ $\rightarrow$ 1$^-$ transitions, displayed respectively in \cref{fig:exp}(e) and (f), with the one for the 1$^-$ $\rightarrow$ $0$ transition in \cref{fig:exp}(d), reveals that the second excess electron occupies the same orbital as the first one (see nodal-plane orientation).
Statistical validation and additional data sets are provided in the SM~\cite{SM}.
The intrinsic molecular Jahn--Teller is though insufficient to explain this result, as explained by the following theoretical analysis.

\emph{Model}. We model the molecule in the AC-STM with the Hamiltonian
\begin{equation}
	\hamil[] = \hamil[mol] + \hamil[tip] + \hamil[tun]+ \hamil[as] + \hamil[vib] + \hamil[e-ph]\,,
\end{equation}
where the minimal electronic description of the molecule (mol), metallic tip (tip) and tunnelling coupling (tun) is complemented by a small environmental asymmetry (as), the set of molecular and substrate vibrational modes (vib), and by their coupling to the electronic degrees of freedom (e-ph).
We write the molecular Hamiltonian in the Fock space of the four frontier orbitals, which are the singly occupied molecular orbital (SOMO, $S$), the highest occupied molecular orbital (HOMO, $H$) and the two degenerate lowest unoccupied molecular orbitals (LUMOs, $L_{xz}$ and $L_{yz}$). 
The latter two are shown in \cref{fig:exp}(b).
For more details on the modelling of the molecule, tip and tunneling part of the Hamiltonian we refer to the SM~\cite{SM} and our previous publications \cite{Sobczyk2012, Donarini2012, Siegert2015a,Siegert2016, Frankerl2021}.
Additionally, we incorporate an environmental asymmetry via $\hamil[as] = -\delta_{\rm as} (\op[xz]{n}-\op[yz]{n})$, with $\delta_{\rm as} = 4k_BT = \SI{2}{\meV}$ into our model.
\begin{figure}
	\centering
	\includegraphics[width=.85\linewidth]{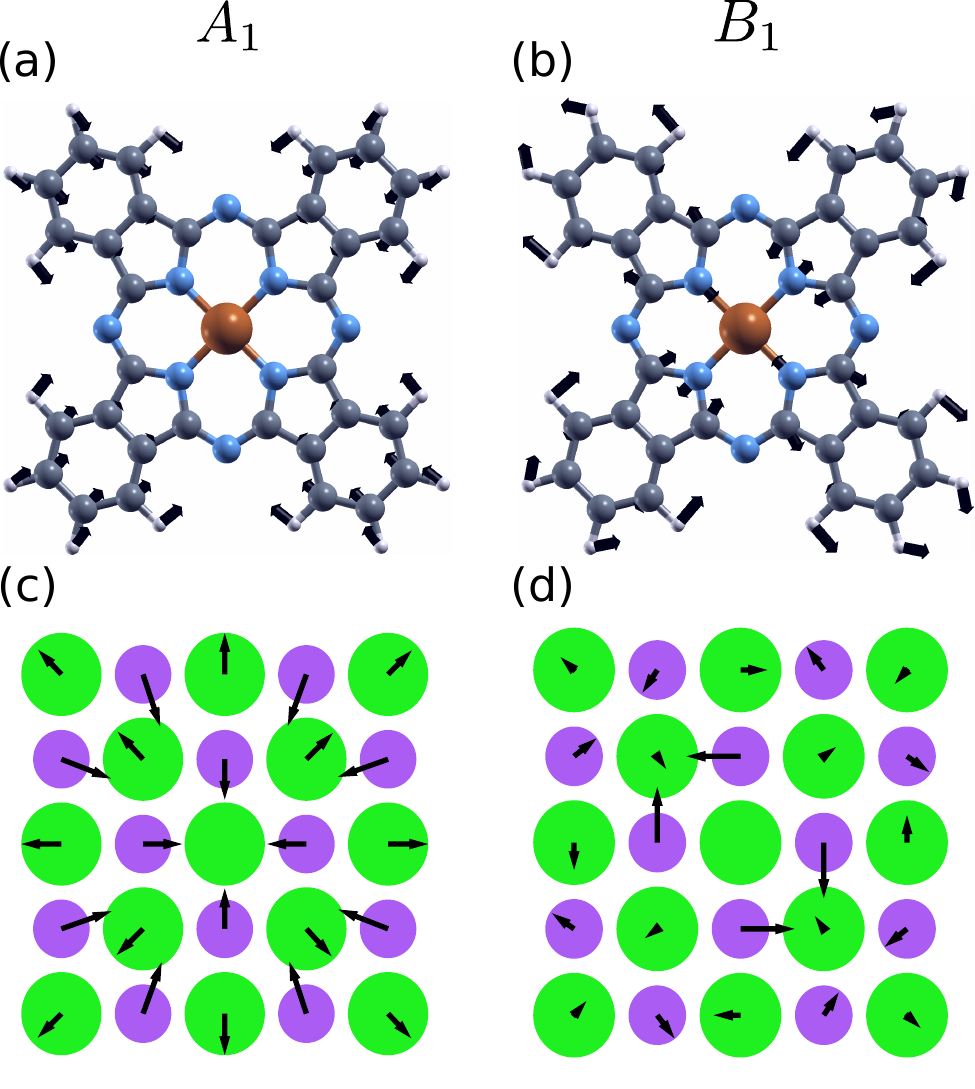}
	\caption{
		(a), (b) Examples of a $A_1$ and a $B_1$ molecular vibrational mode, respectively (Cu in brown, C in grey, N in blue, H in white).
        (c), (d) In-plane displacement of the salt atoms (Cl$^-$ in green, Na$^+$ in purple) corresponding to the $A_1$ ($B_1$) mode, respectively.
	}
	\label{fig:salt}
\end{figure}

The vibronic Hamiltonian is a collection of harmonic oscillators $\hamil[vib] = \sum_\zeta \hbar\omega_\zeta (\op[\zeta]{a}^\dag \op[\zeta]{a} + \nicefrac{1}{2})$ containing molecular and salt modes. Based on symmetry considerations, the vibrational modes of the system fall into three categories: (i) those not interacting with charges in the molecule, (ii) those affecting both LUMOs in the same way, and (iii) those interacting differently with each of the LUMOs.
The $C_{4v}$ symmetry of the combined salt and molecular system facilitates this classification (c.f.~SM~\cite{SM} for more details on the treatment of the modes).
We ignore the modes in the first category.
The ones constituting the second category stabilize charge states in the molecule.
They transform as the $A_1$ irreducible representation.
We call them symmetric (S) since they do not reduce the symmetry of the system.
They are only sensitive to the global charge on the LUMOs, thus
\begin{equation}
	\hamil[e-ph]^{\rm S} = \sum_{\alpha \in \{A_1\}} g_\alpha (\op[xz]{n}  + \op[yz]{n}) \left( \op[\alpha]{a}^\dag + \op[\alpha]{a} \right)\,.
\end{equation}
The modes in the third category affect the charge distribution within a specific charge state. We refer to them as antisymmetric (AS) modes, since they lower the symmetry of the molecule. They belong to the $B_1$ or $B_2$ irreducible representations. Due to energetic considerations (cf. SM~\cite{SM}) we completely neglect the $B_2$ modes. The antisymmetric ($B_1$) modes couple to the imbalance between the populations of the  $L_{xz}$ and $L_{yz}$ orbitals, i.e.
\begin{equation}
	\hamil[e-ph]^{\rm AS} = \sum_{\beta \in \{B_1\}} g_\beta (\op[xz]{n} - \op[yz]{n})\left( \op[\beta]{a}^\dag + \op[\beta]{a} \right)\,.
\end{equation}

We calculated the molecular mode energies as well as the corresponding electron-phonon couplings $g_{\alpha/\beta}$ from first principles using \verb|Inelastica| \cite{Soler2002,Frederiksen2007}, thus obtaining 14 symmetric ($A_1$)  and 14 antisymmetric ($B_1$) modes (see SM for details~\cite{SM}). The coupled salt modes can be combined to obtain just a single symmetric ($A_1$) and a single antisymmetric ($B_1$) mode. We take the corresponding electron-phonon couplings as free parameters in our theory and set them to $g^\mathrm{salt}_{A_1} = \SI{68}{\meV}$ and $g^\mathrm{salt}_{B_1} = \SI{40}{\meV}$, assuming for the energy the one of the transversal optical phonon mode $\hbar\omega_\mathrm{TO} = \SI{20}{\meV}$ \cite{Raunio1970}. A microscopic derivation of this minimal model yielding couplings constants of comparable strength is given in the SM~\cite{SM}. We show a comparison between one symmetric molecular mode and the symmetric salt mode in \cref{fig:salt}(a) and (c). The antisymmetric case is displayed in \cref{fig:salt}(b) and (d). The influence of the symmetric modes on the many-body spectrum is obtained by a canonical Lang--Firsov transformation, leading to a renormalization of the single-particle energies $\tilde{\epsilon}_i = \epsilon_{i} - \sum_\alpha \nicefrac{g_\alpha^2}{\hbar\omega_\alpha}$ as well as of the direct Coulomb interaction terms $\tilde{V}_{ijji} = V_{ijji} - 2\sum_{\alpha} \nicefrac{g_\alpha^2}{\hbar\omega_\alpha}$ and $i,j \in \{ L_{xz}, L_{yz} \}$. Thus, the energy renormalization induced by the symmetric salt mode explains the stabilization of charged molecules and atoms adsorbed on polar dielectrics due to the Franck--Condon blockade \cite{Mitra2004,Koch2006}.     
For antisymmetric modes, the exchange and pair hopping between LUMOs shows an intricate interplay with the electron-phonon coupling. In this case, the Lang--Firsov transformation does not eliminate the coupling between the electronic and the mechanical degrees of freedom, which persists in the transformed exchange and pair-hopping terms. We identify, instead, a single JT active mode and resort for our analysis into a semi-classical adiabatic approximation. To this end, we introduce the canonical displacement operators
\begin{equation}
	\op[m]{Q} = \sum_{\beta \in B_1} \sqrt{\hbar\omega_\beta} A_{m\beta}\left( \op[\beta]{a}^\dag+\op[\beta]{a} \right),
\end{equation}
where the first row of the transformation matrix reads $A_{1\beta} = \nicefrac{1}{g} \sqrt{\nicefrac{g_\beta^2}{\hbar\omega_\beta}}$ and yields the JT active mode $\op[\rm AS]{Q}:= \op[1]{Q}$. The JT distortion is thus described by the operator 
\begin{equation}
	\hamil[JT](Q_{\rm AS}) = \frac{Q_{\rm AS}^2}{4} +  g\,(\op[xz]{n} -  \op[yz]{n}) Q_{\rm AS}\,
\end{equation}
being $Q_{\rm AS}$ the classical displacement of the JT active mode, with the effective coupling $g= \sqrt{\sum_\beta \nicefrac{g_\beta^2}{\hbar\omega_\beta}}$. 
On this basis, we simulate the experimentally investigated many-body eigenstates and energies by diagonalizing the effective Hamiltonian 
\begin{equation}
\hamil[eff] = \op[\rm mol]{{\tilde H}} + \hamil[as] + \hamil[JT] (Q_{\rm AS})\,,
\end{equation}
along the coordinate $Q_{\rm AS}$ of the JT-active antisymmetric mode, with $\op[\rm mol]{{\tilde H}}$ containing the Lang--Firsov renormalization due to the symmetric modes.\\
\begin{figure*}[ht]
	\begin{center}
	\includegraphics[width=\linewidth]{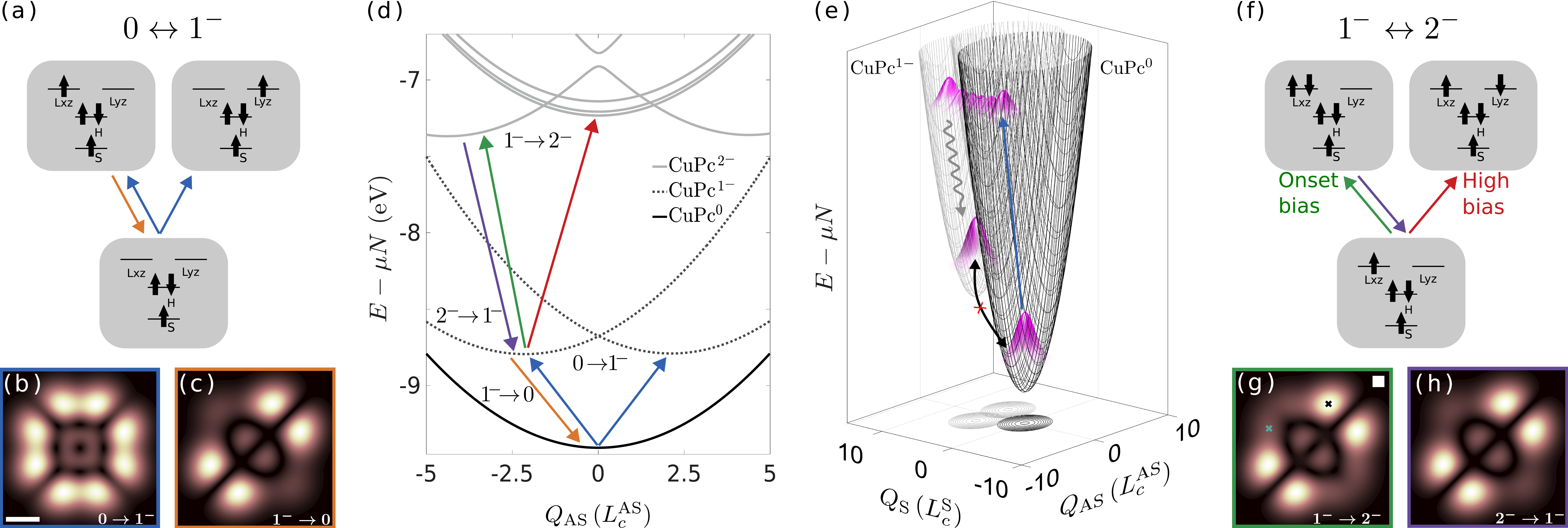}
	\end{center}
	\caption{
        Many-body states, energies and transitions.
		(a) Many-body states relevant for the transitions between neutral and anionic states. 
		(b) $0\rightarrow 1^-$ at $V_b = \SI{1.375}{\V}$, shows 4-fold symmetry. All scale bars are $\SI{5}{\AA}$ and we use a Gaussian broadening of 4\AA.
		(c) $1^-\rightarrow 0$ at $V_b = \SI{0.5}{\V}$, 4-fold symmetry is broken.
	    (d) Adiabatic potential energy surfaces (APESs) as a function of the antisymmetric deformation coordinate ($Q_{\rm AS}$). The arrows represent the transitions to the corresponding ground states at the energy given by APES. 
        (e) The $0\rightarrow 1^-$ transition involves a strong excitation of the symmetric modes ($Q_{\rm S}$ coordinate) due to the large Huang--Rhys factor. The latter strongly suppresses the direct transition between ground states (black arrow). The allowed transition from the ground state to the vibronically excited state (blue arrow) is followed by the relaxation dynamics towards the vibrational ground state (wavy downward arrow).
	    (f) Many-body states relevant for the transitions $1^-\leftrightarrow 2^-$. At a higher bias also the $L_{yz}$ orbital is populated, as indicated by the red arrow.
        (g) $1^- \rightarrow 2^-$ at $V_b = \SI{2.75}{\V}$. At the onset bias the second added electron occupies the same $L_{xz}$ orbital.
	    The crosses indicate the positions reported in \cref{fig:atd_bias} (a).
	    (h) $2^- \rightarrow 1^-$ at $V_b = \SI{1.8}{\V}$.
}
	\label{fig:apes_maps}
\end{figure*}
\emph{Electronic structure}. Some of the eigenenergies of $\hamil[eff]$ are the adiabatic potential energy surfaces (APES) plotted in \cref{fig:apes_maps}. The lowest parabola in \cref{fig:apes_maps}(d) is the APES of the neutral state. This neutral ground state defines the reference configuration ($Q_{\rm AS} = 0$). The anionic states with an excess electron in one of the LUMO levels are orbitally degenerate for $Q_{\rm AS} = 0$. Their energy is lowered upon deforming along $Q_{\rm AS}$. Thus, the anionic ground state has one extra electron occupying the $L_{xz}$ ($L_{yz}$) orbital for $Q_{\rm AS} < 0$ ($Q_{\rm AS} > 0$). The small environmental asymmetry $\delta_{\rm as}$ ensures a global minimum with the excess electron in $L_{xz}$. The singlet-triplet splitting is not resolved on this energy scale. So far, the spectrum reflects the well-known JT effect. 
The richer structure of the dianionic APESs express two qualitatively different distributions of the excess electrons.
For one excess electron in each of the $L_{xz}$ and $L_{yz}$ orbitals, the coupling to the JT mode vanishes. Thus, the lowest energy of these states (with different spin multiplicities) is found at $Q_{\rm AS} = 0$ in the upper part of \cref{fig:apes_maps}(d). On the other hand, states with both excess electrons in the same former LUMO strongly couple to the JT-mode. Their ground state is found at $Q_{\rm AS}\neq 0$, analogously to the anionic configuration, though with an even larger JT deformation and reorganization energy. Pair hopping between the real-valued orbitals induces the avoided crossing at $Q_{\rm AS} = 0$, characteristic of the pseudo JT effect \cite{Bersuker2006}.\\
\indent \emph{Many-body transition rates}. A qualitative understanding of the measured charge-transitions is obtained by combining the electronic structure (detailed derivation in SM~\cite{SM}) with few other constraints imposed by the experimental set-up. 
Specifically, in AC-STM a single electron is tunneled back and forth per cantilever oscillation cycle of 
$\SI{34}{\mu\s}$ duration. Hence, the system thermalizes between consecutive tunneling events.
Moreover, the spectral broadening induced by the symmetric modes is much larger than the environmental asymmetry, which in turn exceeds the temperature.
Those premises imply a 4-fold symmetric neutral to anionic transition; it resembles the (incoherent) superposition of the $L_{xz}$ and $L_{yz}$ orbitals (with additional toggling), as it involves the spectrally unresolved states associated to both local minima with $Q_{\rm AS} >0$ and $Q_{\rm AS} < 0$ [\cref{fig:exp}(c) and \cref{fig:apes_maps}(b)].
In contrast, environmental asymmetry ensures, for the thermalized anion, the predominant occupation of the state with $Q_{\rm AS}<0$; consequently the anionic to neutral transition acquires the 2-fold rotational symmetry of the  $L_{xz}$ orbital [\cref{fig:exp}(d) and \cref{fig:apes_maps}(c)].
Mere energetic arguments would predict also for the transition between the anion and the dianion the participation of the states corresponding to both the deformed local minima in the dianionic APES.
Single electron tunnelling, though, forbids the contribution of the state with $Q_{\rm AS} >0$ and double occupation of the $L_{yz}$ orbital, and ultimately yields a many-body transition resembling again $L_{xz}$ [\cref{fig:exp}(e) and \cref{fig:apes_maps}(g)].
The same appearance is expected for the transition from the dianionic to the anionic ground state [\cref{fig:apes_maps}(h)] although quantum fluctuations in the configuration of the dianion do not completely exclude small contributions from the $L_{yz}$ orbital, cf.~\cref{fig:exp}(f) and the SM~\cite{SM} for dependence of the dianionic ground state on the deformation coordinate.   
Electron tunnelling implies displacements of both symmetric and antisymmetric modes. Their interplay is visualized in \cref{fig:apes_maps}(e). The large electron-phonon coupling to the symmetric modes strongly suppresses the transitions between ground states along the effective symmetric coordinate $Q_{\rm S}$ [black arrow in \cref{fig:apes_maps}(e)]. On the contrary, with respect to the antisymmetric modes transitions between ground states are possible. This phenomenon, visualized for the neutral to anionic transition holds for all other considered transition between states with different charges. The theoretical transition maps in \cref{fig:apes_maps} and \cref{fig:atd_bias} are calculated with Fermi's golden rule as described in the End Matter. 
The tunneling rates are depicted in \cref{fig:apes_maps} as a function of the tip position and show a strong similarity to the experimental results, with also a remarkable agreement between all the experimental and theoretical transition voltages. 

We notice, in this respect, that the $1^-$ to $2^-$ transition denoted with a green arrow in \cref{fig:apes_maps} is calculated, in accordance to the experiment, at the \emph{onset} of the anion to dianion transition ($V_b = \SI{2.75}{\V}$). 
%
\begin{figure}
	\begin{center}
	\includegraphics[width=\linewidth]{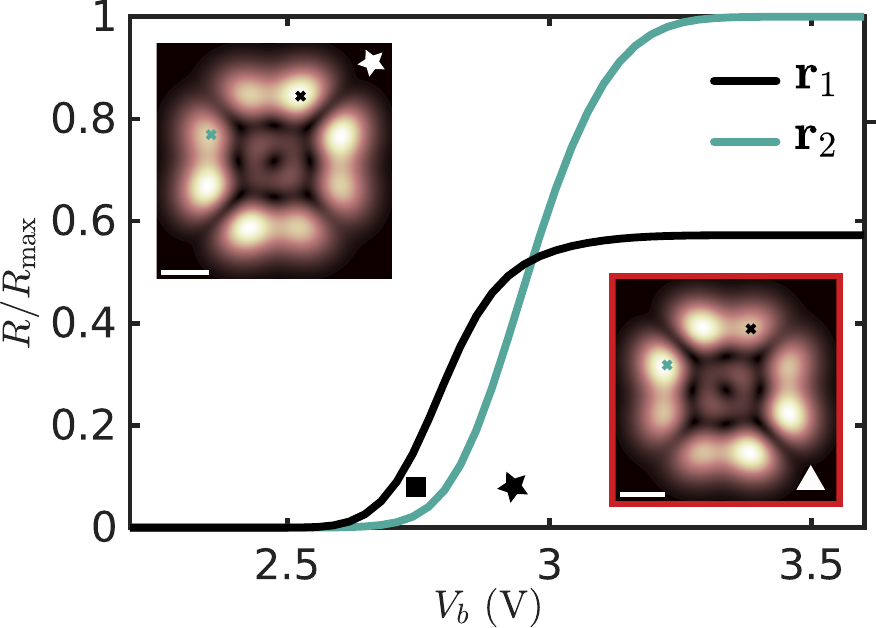}
	\end{center}
	\caption{Bias dependence of the anionic to dianionic transition rate calculated in two different positions. 
 The rates are given as a fraction of the maximum calculated value $R_{\rm max}$.
 The ratio between the  rates varies with the bias, due to the opening of a further transition at higher bias [Fig.~2(f)]. The insets depict maps of the transition rate. The crosses mark the $\bm{r}_1$ and $\bm{r}_2$ positions used in the main figure. Star: Calculated at $V_b = \SI{2.9}{\V}$. Triangle: Map of the transition rate calculated at $V_b = \SI{3.5}{\V}$. At this bias the transition involving the $L_{yz}$-orbital is dominant [Fig.~2(g)].
	 Scale bar is 5 \AA.}
	\label{fig:atd_bias}
\end{figure}
At higher biases we expect to open also transitions to the excited dianionic states, as sketched by the red arrows in \cref{fig:apes_maps}(d) and (f). 
The bias dependence of the transition rate is  shown in \cref{fig:atd_bias}. 
We evaluate it in the tip positions visualized by the black and teal crosses in the insets. They select tunnelling events involving a single orbital, respectively $L_{xz}$ for $\bm{r}_1$ and $L_{yz}$ for $\bm{r}_2$. The form of the dianionic ground state determines the lower bias onset for the rate in position $\bm{r}_1$. At a higher bias voltages both rates saturate, with the one at $\bm{r}_2$ being twice as large as the one at $\bm{r}_1$: the unoccupied orbital offers two spin channels as compared to the single one for an orbital already occupied by one electron. The predicted change of the full rate map is shown in the insets of \cref{fig:atd_bias}. The comparison to \cref{fig:apes_maps}(g) and \cref{fig:exp} shows the emergence of the orthogonal LUMO ($L_{yz}$ in our model) with an overall stronger signal~\footnote{Additional measurements taken during the review process seem to confirm this theoretical prediction.}.
Control over the landscape of the dianionic APES should also be obtained by changing the underlying salt. It is experimentally known that RbI, NaCl, and LiF have increasing influence onto the spectral broadening of molecular electronic states \cite{Repp2010b,Pavlicek2013} and thus on the reorganization energy of the symmetric modes. We hypothesize a similar trend also for the antisymmetric modes, with a possible reversal of the orbital occupation sequence of CuPc on RbI with respect to the one on NaCl and LiF.

\emph{Conclusions}. We have shown how the electron-phonon coupling between a charged molecule and the underlying substrate phonons determines the specific charge configuration within a molecule going beyond mere charge stabilization. We extend to this respect existing observations of charge localization in oligophenylene \cite{Patera2020} and pentacene \cite{HernangomezPerez2020} to systems with JT distortion. Generally, we predict that the salt substrate has a larger effect on molecules with orbital degeneracy. This approach could pave the way to control charge configurations in molecules depending on the used substrate, with potential relevance in the study of intermolecular electron transfer, surface chemical reactivity and charge sensing. We also envisage the interplay between JT distortion of neighbouring molecules for the realization of cellular automata based on molecular arrays in which the information is locally stored in the charge configuration of a single molecule.

\emph{Acknowledgements}.
We thank Luisa Raischl
for the experimental support. Moreover, we gratefully acknowledge financial support from the Deutsche Forschungsgemeinschaft (DFG, German Research Foundation) within Project-ID 314695032 - SFB 1277 and RE2669/6-2, the ERC Synergy Grant MolDAM (no. 951519), the Spanish MCIN/AEI/ 10.13039/501100011033 (PID2020-115406GB-I00 and PID2023-146694NB-I00), the Basque Department of Education (PIBA-2023-1-0021), and the Euskampus Transnational Common Laboratory \emph{QuantumChemPhys}.

\bibliography{library}
\appendix
\section{End Matter}
\label{sec:methods}
\textit{Methods}.
The transition rate from an initial thermalized state towards a vibrationally excited state with neighboring charge reads
\begin{widetext}
\begin{subequations}
\label{eq:Rates}
\begin{align}
	&R^{N\rightarrow N+1} = \sum_{\{n_\alpha\},f} \sum_{i,j,\sigma}\prod_{\alpha \in \{A_1\}}\!\!
 \Gamma_{i\sigma,j\sigma}(\bm{r}_{\rm tip})
	P_{{\lambda_\alpha}}(\{n_\alpha\})f^+\big(E_f - E_g - e\;\! c V_b\big)
  \langle N+1,f  | \op[i\sigma]{d}^\dag | N,g \rangle  \langle N,g | \op[j\sigma]{d} | N+1,f \rangle\,,
	\label{eq:rate_up_vib}\\
	&R^{N+1\rightarrow N} = \sum_{\{n_\alpha\},f} \sum_{i,j,\sigma}\prod_{\alpha \in \{A_1\}}\!\!
 \Gamma_{i\sigma,j\sigma}(\bm{r}_{\rm tip})
	P_{{\lambda_\alpha}}(\{n_\alpha\})f^-\big(E_g - E_f - e \;\! c V_b\big)
  \langle N+1,g | \op[i\sigma]{d}^\dag | N,f \rangle \langle N,f | \op[j\sigma]{d} | N+1,g \rangle\,,
  \label{eq:rate_down_vib}
\end{align}
\end{subequations}
\end{widetext}
where the single-particle rate matrix is given by $\Gamma_{i\sigma,j\sigma}(\bm{r}_{\mathrm{tip}}) = \nicefrac{2\pi}{\hbar} \sum_{k} t_{k,i\sigma}^* t_{k,j\sigma} \delta(E_f - E_g - \epsilon_k)$. The symmetric modes' Franck--Condon factors yield the Poisson distributions $P_{\lambda_\alpha}(\{n_\alpha\})$, which depend on the Huang--Rhys factors $\lambda_\alpha = \nicefrac{g_\alpha}{\hbar\omega_\alpha}$ and $\{n_\alpha\}$ is the set of the final state's vibrational excitations with respect to all modes $\alpha \in \{A_1\}$. Due to the large electron-phonon coupling $\lambda_{\rm salt} \approx 5$, the salt contribution converges to a Gaussian and it is the main source of spectral broadening \cite{Repp2005a}.   
The Fermi function $f^+$ (and related $f^- = 1-f^+$) with the tip temperature $T = \SI{6}{\K}$ and chemical potential $\mu=\SI{-4.65}{\eV}$ is calculated at the difference between the molecular many-body energies shifted by the fraction $c = 0.6$ of sample bias $V_b$ dropping between the molecule and the tip. The energy of the final state $E_f = E_f^e + \sum_\alpha \hbar\omega_\alpha n_\alpha$ include the electronic potential energy $E_f^e$ (the local minima of the APES in \cref{fig:apes_maps}) and the excitation energies of the symmetric modes, while $E_g$ is the ground state energy of the initial state. Finally, \cref{eq:rate_up_vib,eq:rate_down_vib} include the transition matrix elements between different charge states. To this end, we denote the initial state with $\ket{N,g}$ ($\ket{N+1,g}$) and the final state with $\ket{N+1,f}$ ($\ket{N,f}$). The sum over $f$ extends to all possible states accessible from the designated initial state via a single electron tunneling event. 
\onecolumngrid
\newpage
\section{Experimental details}
Experiments were carried out with a low-temperature scanning tunneling and atomic force microscope equipped with a qPlus tuning fork \cite{Giessibl2000} (resonance frequency $f_0 \approx 29.1$ kHz, spring constant $k = 1.8$ kN m$^{-1}$, quality factor $Q \approx 3 \times 10^4$) in ultrahigh vacuum ($p = 2 \times 10^{-10}$ mbar) and at a temperature of 6.2 K. As substrate we used Cu(111) single crystal covered with $>$20ML of NaCl. The molecules were deposited onto the sample kept at $\approx 7$ K inside the microscope. Bias voltages are given as sample bias with respect to the tip. Positive constant-height offsets $\Delta z$ correspond to a distance decrease with respect to the AFM $\Delta f$ set point above the clean NaCl. For AC-STM experiments, the ac voltage pulses were produced by an arbitrary waveform generator (Agilent 33522A) and fed to the microscope head through semirigid-coaxial high-frequency cables.

\section{Side-by-side display of experimental and theoretical transition maps}
\begin{figure}[h!]
    \centering
    \includegraphics[width=0.8\linewidth]{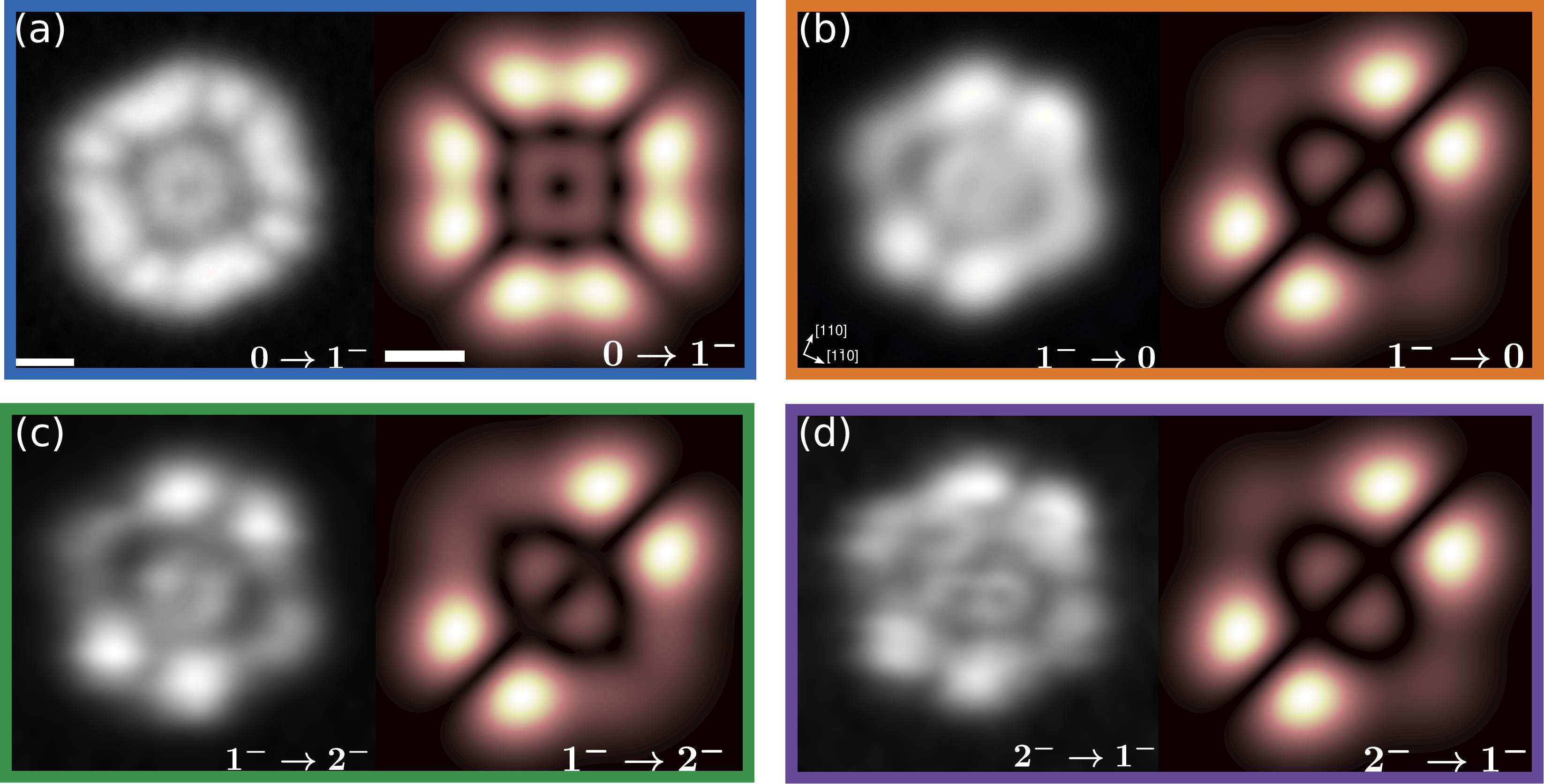}
    \caption{Direct comparison of the experimental (grey) and theoretical (copper) transition maps. Note that the azimuthal orientation of the molecule is not the same in the experimental and theoretical panels; see also main text for further discussion. 
    Transitions from:
    (a) Neutral to anion at $V_{\rm dc} = \SI{1.0}{\V}$, $V_{\rm ac} = 0.75$ V peak-to-peak (V$_{\rm pp}$) (exp) and $V_b = \SI{1.375}{\V}$ (theory).
    (b) Anion to neutral at $V_{\rm dc} = \SI{1.0}{\V}$, $V_{\rm ac} = \SI{1.0}{\V}_{\rm pp}$ (exp) and $V_b = \SI{0.5}{\V}$ (theory).
    (c) Anion to di-anion at $V_{\rm dc} = \SI{2.3}{\V}$,  $V_{\rm ac} = \SI{0.9}{\V}_{\rm pp}$ (exp) and $V_b = \SI{2.75}{\V}$ (theory).
    (d) Di-anion to anion at $V_{\rm dc} = \SI{2.3}{\V} $, $V_{\rm ac} = \SI{1.0}{\V}_{\rm pp}$ (exp) and $V_b = \SI{1.8}{\V}$ (theory).
    Scale bars are 5 \AA.}
    \label{fig:enter-label}
\end{figure}

\section{ Statistical validation of experimental data}
\begin{figure}[h!]
    \centering
    \includegraphics[width=0.8\linewidth]{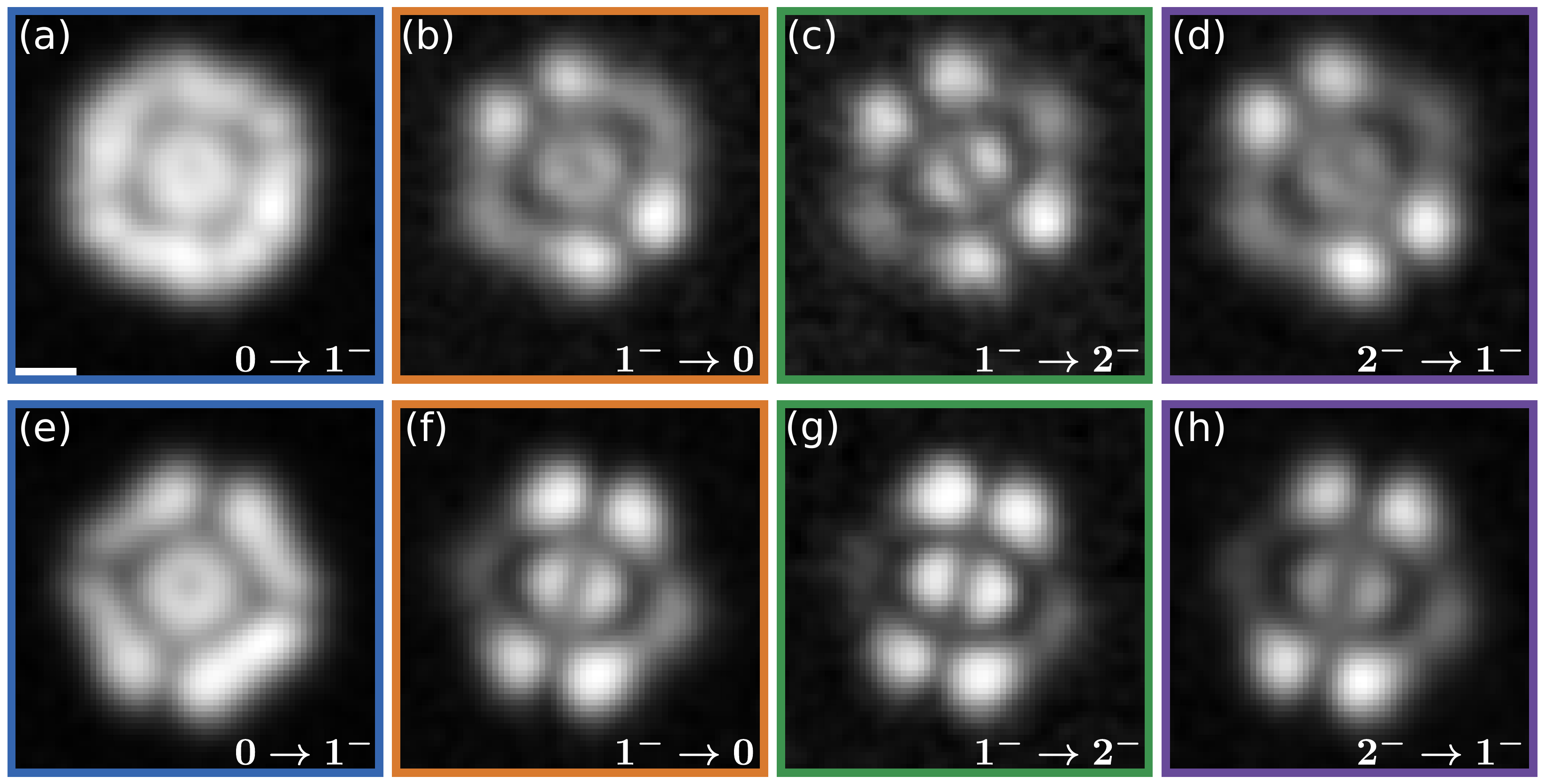}
    \caption{Experimental data sets for two additional molecules. We show one molecule in panels (a)-(d) and the other in panels (e)-(f).
    (a) Neutral to anion at $V_{\rm dc} = \SI{1.125}{\V}$,  $V_{\rm ac} = \SI{0.8}{\V}_{\rm pp}$ and $\Delta z = 0.7$ \AA.
    (b) Anion to neutral at $V_{\rm dc} = \SI{1.125}{\V}$, $V_{\rm ac} = \SI{0.8}{\V}_{\rm pp}$ and $\Delta z = 0.7$ \AA.
    (c) Anion to di-anion at $V_{\rm dc} = \SI{2.505}{\V}$,  $V_{\rm ac} = \SI{0.8}{\V}_{\rm pp}$ and $\Delta z = 1.5$ \AA.
    (d) Di-anion to anion at $V_{\rm dc} = \SI{2.505}{\V} $, $V_{\rm ac} = \SI{0.8}{\V}_{\rm pp}$ and $\Delta z = 1.5$ \AA.
    (e) Neutral to anion at $V_{\rm dc} = \SI{0.795}{\V}$,  $V_{\rm ac} = \SI{0.8}{\V}_{\rm pp}$ and $\Delta z = 0.8$ \AA.
    (f) Anion to neutral at $V_{\rm dc} = \SI{0.795}{\V}$, $V_{\rm ac} = \SI{0.8}{\V}_{\rm pp}$ and $\Delta z = 0.8$ \AA.
    (g) Anion to di-anion at $V_{\rm dc} = \SI{2.195}{\V}$,  $V_{\rm ac} = \SI{0.8}{\V}_{\rm pp}$ and $\Delta z = 1.5$ \AA.
    (h) Di-anion to anion at $V_{\rm dc} = \SI{2.195}{\V} $, $V_{\rm ac} = \SI{0.8}{\V}_{\rm pp}$ and $\Delta z = 1.5$ \AA.
    $\Delta z$ are given with respect to an AFM set point of $\Delta f = -1.2$ Hz at $V_{\rm dc} = 0.001$ V. Scale bar is 5 \AA. A toggling of the molecule can be triggered by the $0 \to 1^-$ transition, depending on the molecular adsorption site: cf. panels (a) and (e).}
    \label{fig:statistics}
\end{figure}

The single-charging transitions ($0 \to 1^-$ and $1^- \to 0$) were measured for 18 individual molecules. 
Two cases were disregarded completely, since the particular molecules were adsorbed directly 
at a step edge of NaCl and a point defect, respectively.
The reduction of the symmetry from C$_{4v}$ to C$_{2v}$ upon single charging ($0 \to 1^-$ compared to the reverse $1^- \to 0$ transition) has been observed in all cases, except for one.
Since the preferential direction is assumed to be dictated by the environmental asymmetry, it is not surprising that once the symmetry breaking is not observed.
For 7 of the above cases all four charging transitions relevant here ($0 \to 1^-$, $1^- \to 0$, $1^- \to 2^-$, and $2^- \to 1^-$) were measured. 
All of these show the localization of the transition $1^- \to 2^-$ on the same orbital as for the $1^- \to 0$ transition.
Two of the latter data sets are exemplarily displayed in Fig.~\ref{fig:statistics}.

\section{Calculating molecular mode parameters}
\label{sec:e_ph_coupling}
The dependence of the electronic wave functions of the molecule on the nuclear positions is obtained by expanding the electronic Hamiltonian around the reference configuration $\bm{R}_0$ up to first order in the displacement coordinate $\bm{q}$. By definition, the first derivative of the ground state energy for the reference electronic state vanishes. However, this is not the case for the each component of the electronic Hamiltonian. 
In terms of the normal modes coordinates, we thus obtain
\begin{equation}
		\hamil[e-ph] = \sum_{l,\alpha}\left(\frac{\partial {\hamil[e]}}{\partial {q_{l,\alpha}}}\right)_{q_{l,\alpha}=0} 
		\sum_\zeta \bm{v}^\zeta_{l,\alpha} \sqrt{\frac{\hbar}{2M_l\omega_\zeta}}\left( \op[\zeta]{a}^\dag + \op[\zeta]{a} \right),
\end{equation}
with $M_l$ the mass of the atom $l$ and $\bm{v}^\zeta$ the displacement vector of the normal mode $\zeta$ with frequency $\omega_\zeta$.
The molecular Hamiltonian  $\hamil[mol]$ presented in the main text corresponds to $\hamil[e](\bm{r},\bm{R}_0)$. The matrix element describing the electron phonon coupling strength is given by
\begin{equation}
	g_{ij}^\zeta = \sum_{l\alpha} \langle i\mid \frac{\partial {\hamil[e]}}{\partial {q_{l\alpha}}} \mid j\rangle \bm{v}^\zeta_{l\alpha}
	\sqrt{\frac{\hbar}{2M_l\omega_\zeta}} .
\end{equation}
These matrix elements can be evaluated by the \verb|Inelastica| python package \cite{Frederiksen2007}.
To evaluate them we first obtained the molecular geometry with the SIESTA DFT implementation \cite{Soler2002}.
We employed a single-zeta plus polarization (SZP) basis set and used the GGA-PBE exchange functional.
A finite-displacement amplitude of 0.02 {\AA} was used.
Moreover, we checked that using a larger Double-Zeta plus Polarization (DZP) basis set the obtained reorganization energies changes by less than 5 per cent for the relevant modes, thus leaving the qualitative picture unchanged. For details about the calculation of the mode energies $\hbar\omega_\zeta$ and electron phonon couplings $g_{ij}^\zeta$ we refer to Ref.~\cite{Frederiksen2007}.

\section{Separation of symmetric and antisymmetric modes}%
\label{sec:Separation of symmetric and antisymmetric modes}
In this section, we aim to demonstrate the contrasting effects of symmetric and antisymmetric modes. We find that the symmetric modes solely lead to an energy renormalization depending on the total number of charges on the molecule. Conversely, the coupling to the antisymmetric modes varies across different orbital occupations, thereby exerting an influence on the charge distribution of the molecular ground state.

Our discussion begins with the component of the Hamiltonian that encompasses the vibronic degrees of freedom as well as the electron-phonon coupling
\begin{equation}
		\label{eq:H_vib}
		\hamil[vib] + \hamil[e-ph] = \sum_{\zeta} \hbar\omega_\zeta\left(\op[\zeta]{a}^\dag\op[\zeta]{a} + \frac{1}{2}\right) + 
		\sum_{\zeta\sigma}\sum_{i,j = L\pm} g^\zeta_{ij} \op[i\sigma]{d}^\dag \op[j\sigma]{d} \left( \op[\zeta]{a}^\dag + \op[\zeta]{a} \right).
\end{equation}
The sum is taken over all contributing modes $\zeta$, the spin $\sigma$, and the degenerate LUMOs $i$ and $j$, as we assume that any excess charge on the molecule to be confined solely to the LUMOs, thus restricting the electron-phonon coupling to these two orbitals.
The electron phonon coupling strength is proportional to the matrix elements with respect to the derivative of the electronic Hamiltonian, see \cref{sec:e_ph_coupling},
\begin{equation}
	\label{eq:coupling_matrix}
	g_{ij}^\zeta \propto\langle i\mid \frac{\partial {\hamil[e]}}{\partial {q_{l\alpha}}} \mid j\rangle.
\end{equation}
The derivative $\frac{\partial {\hamil[e]}}{\partial {q_{l\alpha}}}$ is an irreducible tensor operator which transforms in the same way as the respective normal mode $\zeta$.
The wave functions $i$ and $j$ transform according to the $E_g$ irreducible representation of the $D_{4h}$ point group. Therefore, for the matrix element in \cref{eq:coupling_matrix} to be nonzero we can only use modes $\zeta$ such that their irreducible representation $\Gamma^\zeta$ is contained in the symmetric product
\begin{equation}
	\left[ E_g\otimes E_g \right]_+ = A_{1g} \oplus B_{1g} \oplus B_{2g}.
\end{equation}
The modes transforming according to the $A_{1g}$ irreducible representation couple to the global excess charge on the LUMOs. We call these the symmetric modes. The associated molecular distortion preserves the symmetry of the system. Therefore, they are not contributing to the JT effect. It is important to consider them, as they give a renormalization to the eigenenergies of the molecule depending on the charge state and they lead to a broadening in the transition rates.

On the other hand, the $B_{1g}$ and $B_{2g}$ modes couple to the unbalance between the occupation of the LUMOs. In particular, this imbalance concerns,  for the $B_{1g}$ modes, the real valued basis $L_{xz}$ and $L_{yz}$ shown in Fig.~\ref{fig:orbs}. Contrary to the totally symmetric $A_{1g}$ modes, these modes can lower the symmetry of the molecule. Thus, they are responsible for the JT distortion.

\subsection{Energy renormalization due to symmetric modes}
In this section, we delve into the impact of the symmetric modes. All relevant parameters, which have been calculated as described in \cref{sec:e_ph_coupling}, are shown in \cref{tab:modes}.
The symmetric modes couple diagonally to the LUMOs, i.e.~$i=j$ in \cref{eq:H_vib},
\begin{equation}
		\hamil[vib]^{A} +  \hamil[e-ph]^{A}= \sum_{\alpha \in \{A_{1g}\}} \hbar\omega_\alpha\left(\op[\alpha]{a}^\dag\op[\alpha]{a} + \frac{1}{2}\right) + 
		\sum_{\alpha \in \{A_{1g}\}} g_\alpha \sum_{i\sigma}  \op[i\sigma]{d}^\dag \op[i\sigma]{d} 
        \left( \op[\alpha]{a}^\dag + \op[\alpha]{a} \right).
\end{equation}
All modes couple equally to the occupation of the $L_{xz}$ and $L_{yz}$ orbitals and thus we omit the subscript specifying the LUMO from the coupling strength $g_\alpha$. The sum over $\alpha$ encompasses all modes transforming according to the $A_{1g}$ irreducible representation and $g_\alpha$ is the coupling strength to the LUMOs of mode $\alpha$. We apply the Lang--Firsov transformation to the full Hamiltonian 
\begin{equation}
	 \tilde{\hat{H}} = 
  e^{\hat{S}} 
  \left( \hamil[mol] + \hamil[tip] + \hamil[tun]+ \hamil[as] + \hamil[vib] + \hamil[e-ph] \right) e^{-\hat{S}}\,,
\end{equation}
where the operator 
\begin{equation}
	\hat{S} = \sum_{\alpha\in \{A_{1g}\}} \sum_{i\sigma} \lambda_\alpha \op[i\sigma]{n} 
 \left( \op[\alpha]{a}^\dag-\op[\alpha]{a} \right),
\end{equation}
with $\lambda_\alpha = \tfrac{g_\alpha}{\hbar\omega_\alpha}$. This transforms the annihilation and creation operators on the molecule as
\begin{equation}
\label{eq:tilde_d}
	\tilde{\hat{d}}_{i\sigma} = \op[i\sigma]{d} \op{X}, \quad
	\tilde{\hat{d}}^\dag_{i\sigma} = \op[i\sigma]{d}^\dag \op{X}^\dag.
\end{equation}
with $\op{X} = e^{-\sum_\alpha \lambda_\alpha (\op[\alpha]{b}^\dag - \op[\alpha]{b})}$.
The phonon operators are shifted by
\begin{equation}
		\tilde{\hat{a}}_\alpha = \op[\alpha]{a} - \lambda_\alpha \sum_{i\sigma} \op[i\sigma]{n}, \quad
		\tilde{\hat{a}}^\dag_\alpha = \op[\alpha]{a}^\dag - \lambda_\alpha \sum_{i\sigma} \op[i\sigma]{n}.
\end{equation}
The transformed molecular Hamiltonian thus obtains a renormalization of the single-particle LUMO energies $\tilde{\epsilon}_{i}  =  \epsilon_{i} - \sum_{\alpha \in A_{1g}} \frac{g^2_\alpha}{\hbar\omega_\alpha}$, and also of the direct Coulomb interaction between the LUMOs $\tilde{V}_{ijji} = V_{ijji} - 2\sum_{\alpha \in A_{1g}} \frac{g^2_\alpha}{\hbar\omega_\alpha}$, with $i,j = L_{xz}\,, L_{yz}$.
Furthermore, the symmetric electron-phonon coupling term vanishes, leaving only the harmonic vibrations in the Hamiltonian which describes the effect of the symmetric modes.
The renormalization of the single-particle LUMO energies and the corresponding Coulomb interaction terms leads to a stabilization of charged states on the molecule.
The Lang--Firsov transformation of the tunneling Hamiltonian yields the $\hat{X}$ operator introduced in Eq.~\eqref{eq:tilde_d}. Its matrix elements in the harmonic oscillator eigenbasis are the Franck--Condon factors giving the Poisson distribution in the rates of Eqs.~(A1) of the main text. Ultimately, the $\hat{X}$ operator and thus the coupling to the $A_{1g}$ modes is responsible for the broadening of the calculated transition rates.

\begin{table}[htpb]
\caption{Reorganization energy, dimensionless coupling $\lambda$ and energy quantum of the $A_{1g}$, $B_{1g}$ and $B_{2g}$ molecular modes. The reorganization energy is obtained as $\Delta E_{\rm reorg} = \hbar\omega \lambda^2$, with $\lambda = g/\hbar\omega.$ The vibrational energy and the reorganization energy are given in meV.}
\label{tab:modes}

\centering

    \begin{tabular}{lcccc||lcccc||lcccc}
        & & $A_{1g}$ modes & & &
        & & $B_{1g}$ modes & & &
        & & $B_{2g}$ modes & & \\   
        \hline
        $\Delta E_\mathrm{reorg}$ & &$\quad \lambda \quad$ & &$\hbar\omega$ &
        $\Delta E_\mathrm{reorg}$ & &$\quad \lambda \quad$ & &$\hbar\omega$ &
        $\Delta E_\mathrm{reorg}$ & &$\quad \lambda \quad$ & &$\hbar\omega$\\ 
        \hline
        61.0   & & 1.435   & & 29.6  &
        28.0   & & 0.389   & & 185.0 & 
        11.8   & & 0.292   & & 148.0 \\
        25.5   & &  0.384  & & 173.3 &
        9.6    & & 0.235   & & 172.7 &
        3.0    & & 0.239   & & 53.7  \\
        21.0   & & 0.522   & & 76.7 &
        6.1    & & 0.269   & & 84.9 &
        2.9    & & 0.130   & & 172.5 \\
        16.2   & &0.406    & & 98.4 &
        1.1    & & 0.085   & & 159.1 &
        2.3   & & 0.132 & & 134.1\\
        8.3& & 0.222& &167.3 &
        0.838  & & 0.078   & & 138.2 &
        1.4   & & 0.108 & & 120.4\\
        2.6& & 0.126& &164.1 &
          0.78   & & 0.080   & & 120.3 &
        1.2   & & 0.108 & & 106.2\\
        2.2& & 0.109& &186.5 &
        0.38   & & 0.047   & & 169.3 &
        1.0   & & 0.298 & & 11.2 \\
        1.3& & 0.080& &195.8 &
        0.34   & & 0.056   & & 110.5 &
        0.7   & & 0.061 & & 178.2\\
        0.5& & 0.059& &135.2 &
        0.29   & & 0.12    & & 19.4 &
         0.4   & & 0.048 & & 197.0\\
        0.5& & 0.062& &120.5 &
        0.10   & & 0.034   & & 90.9 &
        0.2   & & 0.093 & & 25.4 \\
        0.02& & 0.012& &110.5 &
        0.046  & & 0.027   & & 62.7 &
        0.03  & & 0.020 & & 76.7 \\
        0.007& & 0.005& &359.7 &
        0.051  & & 0.0038  & & 359.7 &
        0.009 & & 0.005 & & 358.6\\
        0.006& & 0.010& &67.0 &
        0.0024 & & 0.0035  & & 195.4 &
        0.005 & & 0.006 & & 117.1 \\
        0.004& & 0.003& &358.9 &
        0.0001 & & 0.0006  & & 359.0 &
        0.002 & & 0.003 & & 359.04\\
    \end{tabular}

\end{table}

\subsection{Treatment of antisymmetric modes}
The antisymmetric modes are more sensitive to the actual distribution of the charge between the LUMOs, rather then to their global occupation. The electron vibron coupling Hamiltonian, written for the real basis depicted in Fig.~\ref{fig:orbs}, reads 
\begin{equation}
	\hamil[e-ph]^{B} = 
	\sum_{\beta \in \{B_{1g}\}} g_\beta 
    \left( \op[xz]{n} - \op[yz]{n} \right)
    \left( \op[\beta]{a}^\dag + \op[\beta]{a} \right)
+ 
	\sum_{\beta \in \{B_{2g}\}} g_\beta 
    \left(\op[xz]{d}^\dagger\op[yz]{d} + \op[yz]{d}^\dagger\op[xz]{d}\right)
    \left( \op[\beta]{a}^\dag + \op[\beta]{a} \right),
\end{equation}
where $\beta$ encompasses all modes transforming according to either $B_{1g}$ or $B_{2g}$.
This Hamiltonian gives rise to the Jahn--Teller (JT) effect, since each deformation favors the occupation of a determined LUMO thus lifting their degeneracy. The more complex form of the electron-phonon coupling combined with the structure of the exchange Hamiltonian makes the Lang--Firsov transformation unuseful. The latter does not decouple the electronic from the vibronic degrees of freedom in this case. Therefore, we treat the antisymmetric displacements classically and estimate the ground-state conformation by searching for the minima of adiabatic potential energy surfaces (APES). To this end, we introduce canonical displacements and momenta as follows
\begin{equation}
	\hat{x}_\beta = \sqrt{\hbar\omega_\beta} \left( \op[\beta]{a}^\dag+\op[\beta]{a} \right) ,\quad
	\hat{p}_\beta = \frac{i\hbar}{2\sqrt{\hbar\omega_\beta}} \left( \op[\beta]{a}^\dag - \op[\beta]{a} \right) .
\end{equation}
This transformation allows us to express the Hamiltonian associated to the modes with the $B_{1g}$ and $B_{2g}$ symmetry as
\begin{equation}
	\label{eq:H_JTE}
		\hamil[vib]^B +  \hamil[e-ph]^B =  
  \sum_{\beta \in \{B_{1/2 g}\}} \left( \omega_\beta^2 \hat{p}^2_\beta + \frac{1}{4} \hat{x}_\beta^2 \right) 
  + 
	\sum_{\beta \in \{B_{1 g}\} }\frac{g_\beta}{\sqrt{\hbar\omega_\beta}} 
    \left( \op[xz]{n} - \op[yz]{n} \right)
    \hat{x}_\beta
 +
 \sum_{\beta \in \{B_{2 g}\} }\frac{g_\beta}{\sqrt{\hbar\omega_\beta}} 
    \left(\op[xz]{d}^\dagger\op[yz]{d} + \op[yz]{d}^\dagger\op[xz]{d}\right)
    \hat{x}_\beta.
\end{equation}
We further simplify the problem by ignoring the kinetic terms and by turning $\hat{x}_\beta$ into the classical coordinate  $x_\beta$. At this point we should diagonalize the effective Hamiltonian 

\begin{equation}
\label{eq:H_eff}
\hamil[eff](\{x_\beta\}) = \tilde{\hat{H}}_{\rm mol} + \hamil[as] + \sum_\beta \frac{x_\beta^2}{4} +  \hamil[e-ph]^B(\{x_\beta\})
\end{equation}
and look for the ground state energy and configuration for each charge sector. Given the presence of 14 $B_{1g}$ and 14 $B_{2g}$ modes, determining the energy minimum of this potentials proves impractical. A first screening on the modes can be performed using their reorganization energy $\Delta E_{\rm reorg} \equiv \hbar\omega \lambda^2$. As can be seen from \cref{tab:modes}, the $B_{1g}$ modes provide a higher reorganization energy than the $B_{2g}$ modes. Therefore, we neglect the $B_{2g}$ modes and focus our attention solely on the $B_{1g}$ modes.

\section{Many-body eigenstates}%
\label{sec:Many-body states}
\begin{figure}
\begin{center}
	\includegraphics[width=\linewidth]{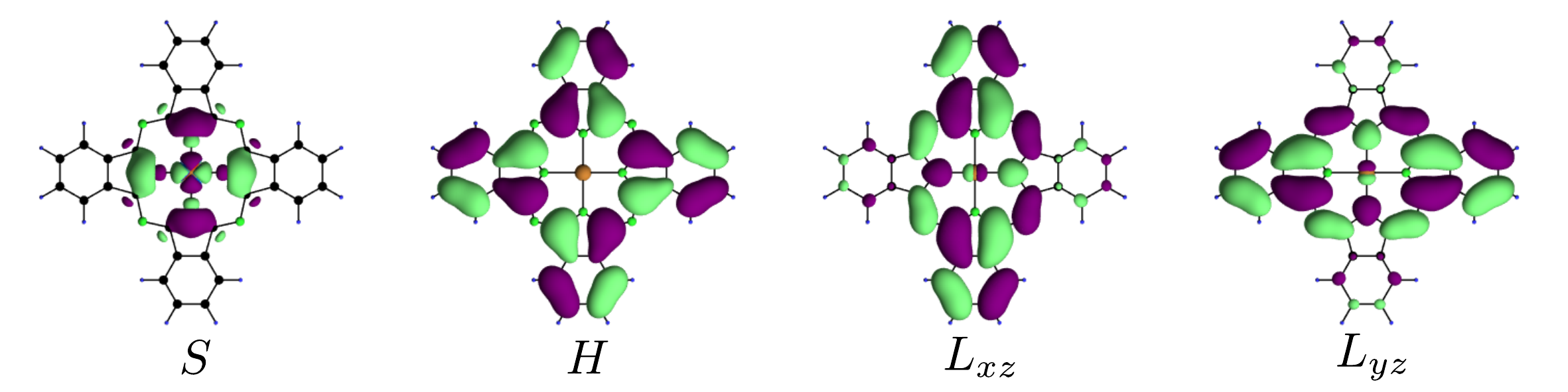}
\end{center}
\caption{Frontier molecular orbitals of CuPc. We show the SOMO ($S$), HOMO ($H$) and the LUMOs in their real valued representation $L_{xz}$ and $L_{yz}$.}
\label{fig:orbs}
\end{figure}

A crucial role for the understanding of the experiments is played by the effective Hamiltonian in Eq.~\eqref{eq:H_eff} from which the low-energy many-body eigenstates of CuPc in its neutral, anionic and dianionic configuration can be extracted. We restrict ourselves to the Fock space generated from the single-particle basis of four frontier orbitals: i.e. the SOMO ($S$), HOMO ($H$) and the two degenerate LUMOs. In particular, we will only consider, in the following, the neutral anionic and dianionic states with, respectively, 0, 1 and 2 electrons in the LUMOs. This further simplification does not affect significantly the low-energy eigenstates of the neutral and anionic molecule, as obtained by comparison with numerical diagonalization performed in the full Fock space associated to the four frontier orbitals. In the dianionic case this approximation excludes a pair of doublets with triple occupation of the LUMOs. However, the latter are not relevant to explain the experimental results, as these states cannot be accessed with a single electron tunneling event from the anionic ground state.  

As the LUMOs are degenerate, they allow for different representations. We show in \cref{fig:orbs} the frontier orbitals with the LUMOs in their real-valued representation ($L_{xz}$, $L_{yz}$). However, in the discussion of the many-body eigenstates, we will also use their rotationally invariant representation which is given by \cite{Siegert2016}
\begin{equation}
		\ket{L+} = \frac{1}{\sqrt{2}} \left( \ket{L_{xz}} + i \ket{L_{yz}} \right), \quad
		\ket{L-} = \frac{1}{\sqrt{2}} \left( \ket{L_{xz}} - i \ket{L_{yz}} \right).
\end{equation}
The Hamiltonian we discuss is $ \hamil[eff](x_1) = \tilde{\hat{H}}_{\rm mol} + x_1^2/4 + \hamil[e-ph]^B(x_1)$, where, to start, we restrict ourselves to the $B_{1g}$ mode with the highest reorganization energy. The renormalization due only to the symmetric modes introduces the same shift to all eigenenergies of a given particle number. In practice, we have
\begin{equation}
	\label{eq:h_eff_1mode}
	\hamil[eff](x_1) =  
		\sum_i (\tilde{\varepsilon}_i + \Delta) \op[i]{n} + \frac{1}{2}\sum_{ijkl}\sum_{\sigma\sigma'}
		\tilde{V}_{ijkl} \op[i\sigma]{d}^\dag \op[k\sigma']{d}^\dag\op[l\sigma']{d}\op[j\sigma]{d}
		+  \frac{x_1^2}{4}
	+ \sum_{\sigma} \frac{g_1}{\sqrt{\hbar\omega_1}}\left( \op[L+\sigma]{d}^\dag \op[L-\sigma]{d} + \op[L-\sigma]{d}^\dag \op[L+\sigma]{d} \right) x_1\,.
\end{equation}
The summation indices $i$, $j$, $k$ and $l$ run over all four frontier orbitals, i.e.~$i$,$j$,$k$,$l$ = $S$,$H$,$L_{+}$,$L_{-}$. The model comprises the renormalized single particle energies $\tilde{\varepsilon}_i = \varepsilon_i - \sum_\alpha g_\alpha^2/\hbar\omega_\alpha$  with $\varepsilon_S = \SI{-12.0}{\eV}$, $\varepsilon_H = \SI{-11.7}{\eV}$, $\varepsilon_{L+/L-} = \SI{-10.7}{\eV}$, also shifted by a crystal field correction of $\Delta = \SI{2.44}{\eV}$. The Coulomb parameters concerning the direct interaction are also renormalized by the reorganization energy of the symmetric modes, i.e. $\tilde{V}_{ijkl} = V_{ijkl} - 2\sum_\alpha g_\alpha^2/\hbar\omega_\alpha$. Exchange and pair-hopping contributions are instead not influenced by the Lang--Firsov transformation. The bare Coulomb parameters, calculated by direct integration from the frontier orbitals \cite{Siegert2016}, are given in \cref{tab:coulomb}.
The energy of the considered mode is $\hbar\omega_1 = \SI{185}{\meV}$ and the coupling strength $g_1 = \SI{72}{\meV}$.

\begin{table}
    \centering
\begin{tabular}[c]{l r  l r}
    \toprule
    $U_\mathrm{S}$ & 11.352 eV&$J^{\mathrm{ex}}_{\mathrm{HL}}=-\tilde{J}^{\rm p}_{\mathrm{H+-}}$&548 meV\\
    \addlinespace
    $U_\mathrm{H}$ & 1.752 eV & $J^{\mathrm{ex}}_{+-}$  &  258  meV\\
    \addlinespace
    $U_{\rm L} = U_\mathrm{\pm} = U_{+-}$ & 1.808 eV & $J^P_{+-}$  &  168 meV\\
    \addlinespace
    $U_\mathrm{SH}$ & 1.777 eV & $J^{\mathrm{ex}}_{\mathrm{SL}}=-\tilde{J}^{\rm p}_{\mathrm{S+-}}$&9 meV\\
    \addlinespace
    $U_\mathrm{SL}$ & 1.993 eV & $J^{\mathrm{ex}}_{\mathrm{SH}} = J^{\rm p}_{\mathrm{SH}}$  &  2 meV\\
    \addlinespace
    $U_\mathrm{HL}$ & 1.758 eV &   &  \\
    \bottomrule
\end{tabular}
\caption{Coulomb integrals between the frontier orbitals. We indicate with $U_i = V_{iiii}$ the Hubbard energy of the orbital $i$, with $U_{ij} = V_{iijj}$  and $J^{\mathrm{ex}}_{ij} = V_{ijji}$ the direct Coulomb integral and the exchange between the orbitals $i$ and $j$. Finally, $J^{\mathrm{p}}_{ij} = V_{ijij}$ and $\tilde{J}^{\mathrm{p}}_{ijk} = V_{ijik}$ refer to the pair hopping and the split pair hopping, respectively.  Taken from Ref.~\cite{Siegert2015a}.}
\label{tab:coulomb}
\end{table}

\subsection{The neutral molecule}
\begin{figure}
\begin{center}
	\includegraphics[width=0.7\linewidth]{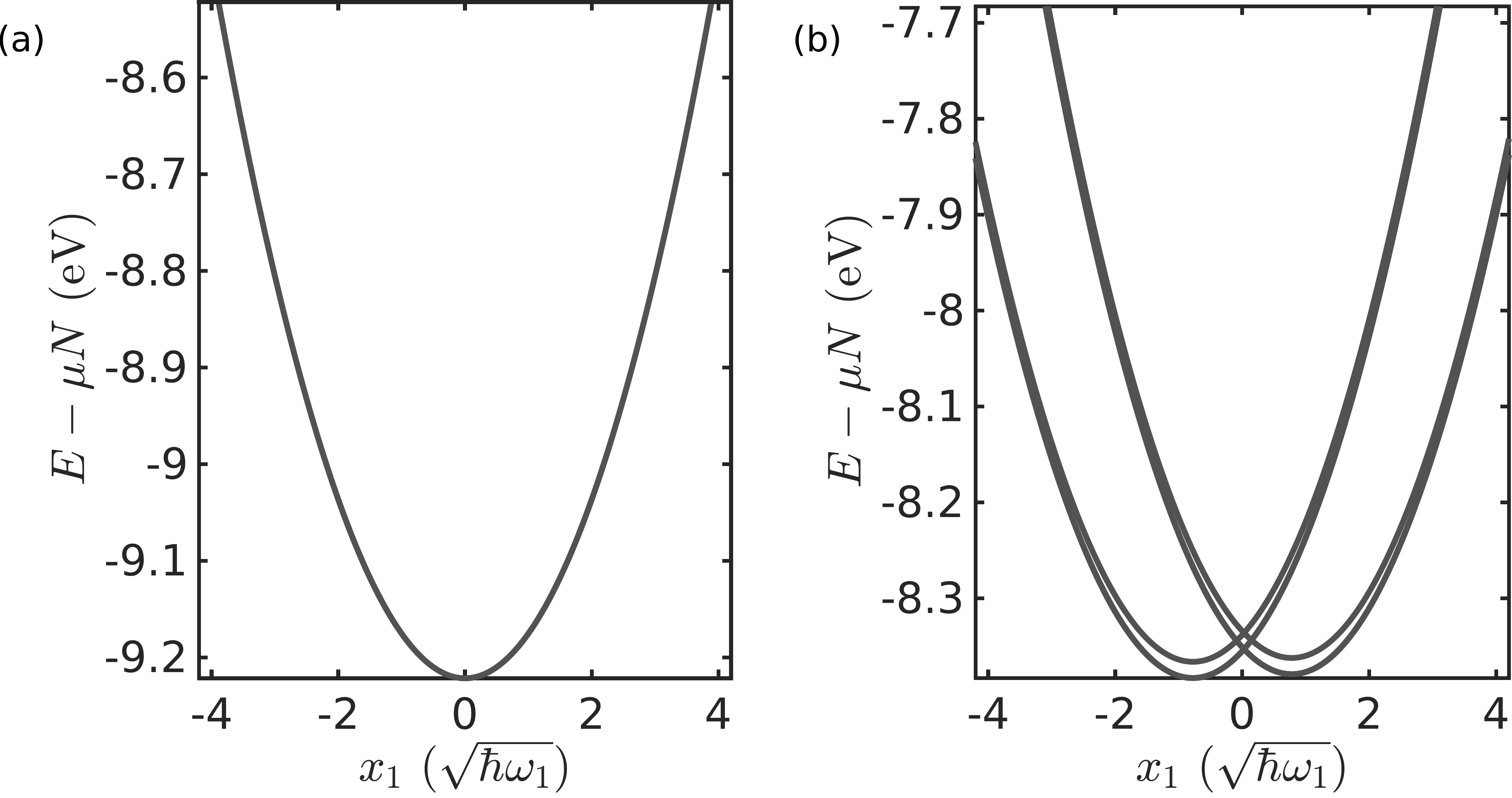}
\end{center}
	\caption{Adiabatic potential energy surfaces along the deformation coordinate belonging to the $B_{1g}$ mode with the highest reorganization energy. The renormalization of the single particle and direct Coulomb interaction due to the $A_{1g}$ modes is taken into account and the chemical potential $\mu = -4.65$ eV. (a) The potential energy of the neutral state is a parabola with its minimum in the reference configuration ($x_1 = 0$), as the electron-phonon coupling does not contribute due to the empty LUMOs. (b) The anionic potential energy is not minimal in the reference configuration. Either the triplet $\ket{4,T_0^{S_z},xz}$  or the triplet $\ket{4,T_0^{S_z},yz}$ are the ground state depending on the sign of the  deformation coordinate in equilibrium.}
\label{fig:ne_an_apes}
\end{figure}

We start our discussion with the eigenstates from the neutral molecule. By diagonalizing \cref{eq:h_eff_1mode} in the reference position $\bm{R}_0$, i.e.~$x_1 = 0$, we find a spin degenerate doublet
\begin{equation}
\ket{3,D_0^\uparrow} = \op[S\uparrow]{d}^\dag\ket{\Omega}, \quad
    \ket{3, D_0^\downarrow} = \op[S\downarrow]{d}^\dag\ket{\Omega},
\end{equation}
where $\ket{\Omega}$ is the reference state with doubly occupied HOMO but empty SOMO and LUMOs. We observe the characteristic unpaired spin in the SOMO, due to its localization and correspondingly high charging energy. The first-excited neutral energy level lies about $\SI{0.8}{\eV}$ above the ground state \cite{Siegert2016} and thus we do not consider it in our calculations.  
The lowest eigenvalue of the Hamiltonian \cref{eq:h_eff_1mode} is  depicted in \cref{fig:ne_an_apes} (a) as  a function of the displacement coordinate $x_1$ given in units of $\sqrt{\hbar\omega_1}$ and represents the adiabatic potential energy surface (APES). As the LUMOs remain unoccupied in the neutral ground state, the electron-phonon coupling does not contribute to the potential energy and we only see the harmonic contribution due to the first antisymmetric mode.

\subsection{The anionic molecule}
The effective Fock space for the CuPc anion comprises the eight states given by $\op[S\sigma]{d}^\dag \op[Li\sigma']{d}^\dag \ket{\Omega}$ with $\sigma,\,\sigma' = \uparrow,\,\downarrow$ and $i = \pm$, which are degenerate in the reference configuration as far as it concerns the direct Coulomb interaction. The exchange coupling between the electrons occupying the SOMO and the LUMO splits this eightfold degeneracy. The eigenstates in the reference configuration can thus be organized as two sets of triplets and singlets
\begin{equation}
\label{eq:basis}
    \begin{aligned}
    &\ket{4,T_0^{+1},i} = \op[S\uparrow]{d}^\dag\op[Li\uparrow]{d}^\dag\ket{\Omega},\\
    &\ket{4,T_0^0,i} = \frac{1}{\sqrt{2}}
    \left(
    \op[S\uparrow]{d}^\dag\op[Li\downarrow]{d}^\dag
    -
    \op[Li\uparrow]{d}^\dag \op[S\downarrow]{d}^\dag
    \right)
    \ket{\Omega},\\
    &\ket{4,T_0^{-1},i} = \op[S\downarrow]{d}^\dag\op[Li\downarrow]{d}^\dag\ket{\Omega},\\
    &\ket{4,S_1,i} = \frac{1}{\sqrt{2}}
    \left(
    \op[S\uparrow]{d}^\dag\op[Li\downarrow]{d}^\dag
    +
    \op[Li\uparrow]{d}^\dag \op[S\downarrow]{d}^\dag 
    \right)
    \ket{\Omega},
    \end{aligned}
\end{equation}
with the triplet orbitally degenerate ground states separated from the singlets by $2 J_{SL} = 18$ meV. These two sets, labeled by $i = \pm$, arise due to the degenerate LUMOs and are mixed by the coupling to the $B_{1g}$ vibrational mode. First, we discuss the two triplet states with $S_z = +1$. The eigenstates of the Hamiltonian \cref{eq:h_eff_1mode} when $x_1\neq 0$ read
\begin{equation}
	\ket{4,T_0^{+1},{xz}} = \frac{1}{\sqrt{2}} \left(\ket{4,T_0^{+1},+} + \ket{4,T_0^{+1},-}\right),\quad
	\ket{4,T_0^{+1},{yz}} = \frac{i}{\sqrt{2}} \left(\ket{4,T_0^{+1},+} - \ket{4,T_0^{+1},-}\right).
\end{equation}
We notice that these states are also eigenstates of \cref{eq:h_eff_1mode} if $x_1 = 0$, thus giving a single set of eigenstates at all deformations, although the corresponding eigenenergies vary as function of $x_1$, as shown in Fig.~\ref{fig:ne_an_apes}. From the definition of the rotationally invariant representation given in Eq.~\eqref{eq:basis} we can express these anionic eigenstates states in the real valued representation as 
\begin{equation}
	\ket{4,T_{0}^{+1},xz} = \op[S\uparrow]{d}^\dag \op[L_{xz}\uparrow]{d}^\dag\ket{\Omega},\quad
	\ket{4,T_{0}^{+1},yz} = \op[S\uparrow]{d}^\dag \op[L_{yz}\uparrow]{d}^\dag\ket{\Omega}.
\end{equation}
The same procedure can be applied to the other 4 triplet states. We obtain
\begin{equation}
	\begin{aligned}
		&\ket{4,T^0_0,iz} = \frac{1}{\sqrt{2}}
  \left(
  \op[S\uparrow]{d}^\dag\op[L_{iz}\downarrow]{d}^\dag
  -
  \op[L_{iz}\uparrow]{d}^\dag\op[S\downarrow]{d}^\dag
  \right)
  \ket{\Omega},\\
		&\ket{4,T^{-1}_0,iz} = \op[S\downarrow]{d}^\dag\op[L_{iz}\downarrow]{d}^\dag\ket{\Omega},
	\end{aligned}
\end{equation}
with $i = x/y$.
Analogously, for the singlet states we obtain, in the real valued basis,
\begin{equation}
	\ket{4,S_1,iz} = \frac{1}{\sqrt{2}} 
 \left( 
 \op[S\uparrow]{d}^\dag\op[L_{iz}\downarrow]{d}^\dag 
 +
 \op[L_{iz}\uparrow]{d}^\dag \op[S\downarrow]{d}^\dag
 \right) \ket{\Omega},
\end{equation}
again with $i =x/y$. In each of the two dimensional subspaces with a given spin $S$ and $S_z$ the Jahn--Teller Hamiltonian in Eq.~\eqref{eq:h_eff_1mode} acquires the canonical form 
\begin{equation}
{\bm H}^{\rm 1-}_{\rm JT} = \frac{x_1^2}{4} {\bf 1}_2 + \frac{g_1}{\sqrt{\hbar\omega}_1}x_1 \boldsymbol{\sigma}_x 
\end{equation}
typical of a $E \otimes b_1$ Jahn--Teller problem \cite{Bersuker2006}.
The sign of the displacement along the $x_1$ coordinate determines whether the triplet-singlet set with the added electron in the $L_{xz}$ or in the $L_{yz}$ orbital experience a decrease or increase in energy. This leads to a preference of one set of triplet-singlet states over the other depending on the direction of distortion. The APES depicted in \cref{fig:ne_an_apes} (b) illustrates this phenomenon. The plot shows four parabolas representing the potential energy of both triplet-singlet sets.

\subsection{The dianionic molecule}

\begin{figure}
\begin{center}
	\includegraphics[width=0.8\linewidth]{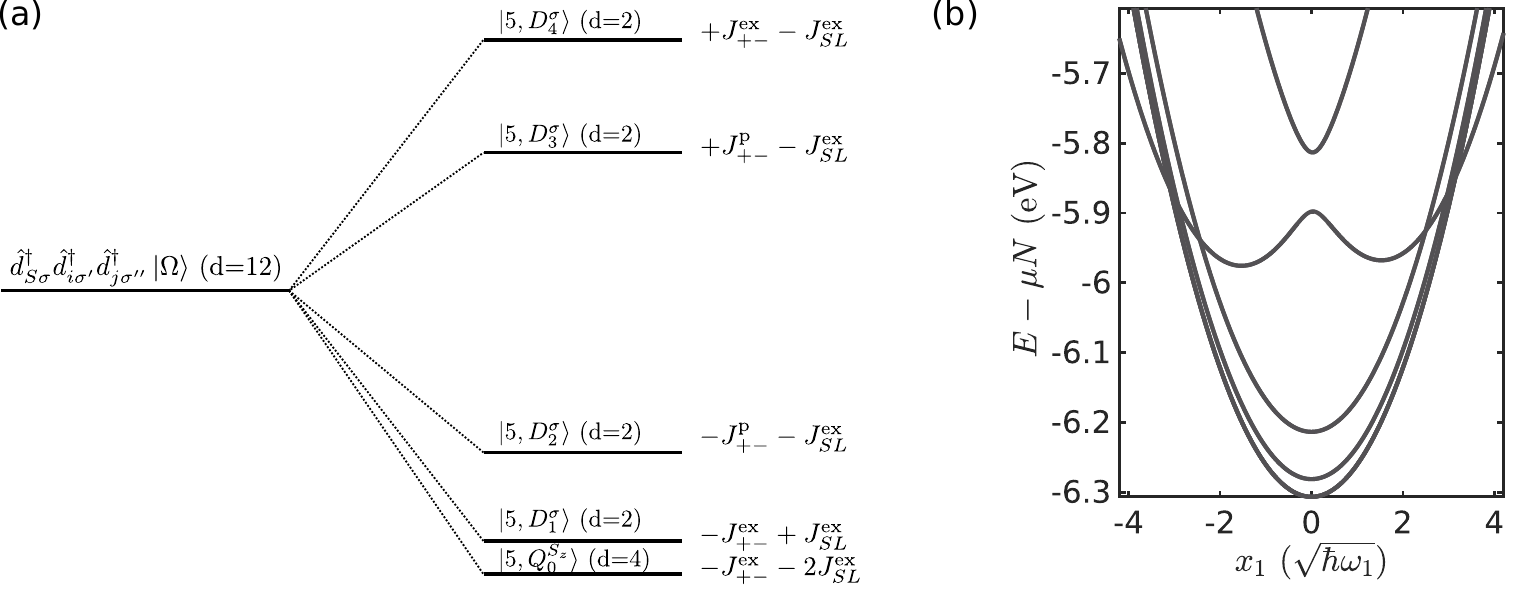}
\end{center}
	\caption{ (a) Scheme of the exchange interaction's influence on the dianionic spectrum at $x_1=0$. 
		The 12-fold degeneracy gets lifted by including the exchange interaction. 
		The states $\ket{5,Q_0^{S_z}}$, $\ket{5,D_1^{\sigma}}$ and $\ket{5,D_2^{\sigma}}$ correspond to the three low lying parabolas in panel (b).
		On the contrary, the states $\ket{5,D_3^{\sigma}}$ and $\ket{5,D_4^{\sigma}}$ are affected by the JT distortion and exhibit an anti-crossing in (b).
		(b) APES of the dianionic molecule. There are 3 parabolas which have their minimum in the $x_1 = 0$ reference configuration. They correspond to the 8 states with the two added electrons distributed between the $L_{xz}$ and $L_{yz}$ orbitals.
	The two lines exhibiting an anti-crossing are describing the energy of the states with the electrons located in one of the two orbitals.
One mode is not enough to lower their energies enough to become the ground state. }
\label{fig:apes_di}
\end{figure}

The effective Fock space for the dianionic molecule consists of twelve states, which, in the rotationally invariant basis, are represented as
\begin{equation}
	\op[S\sigma]{d}^\dag\op[Li\sigma']{d}^\dag\op[Lj\sigma'']{d}^\dag\ket{\Omega},
\end{equation}
where $i,j = \pm$, and $\sigma' = \bar{\sigma}''$ if $i = j$ due to the Pauli exclusion principle. These states are degenerate for what concerns the direct Coulomb interaction (terms involving $U$ in $\hamil[mol]$) with an energy $E_5 = \tilde{\varepsilon}_S + 2(\tilde{\varepsilon}_H + \tilde{\varepsilon}_L) + 5\Delta + \tilde{U}_H + \tilde{U}_L + 2(\tilde{U}_{SH} + \tilde{U}_{SL})+ 4\tilde{U}_{HL}$. By considering the exchange and pair hopping interactions (terms involving $J$ in $\hamil[mol]$), this degeneracy is lifted. We first discuss the spectrum and the eigenstates in the reference configuration $x_1 = 0$. The eigenenergies are given with respect to the degenerate energy.
The scheme of the energy levels is given in \cref{fig:apes_di} (a). The ground state is a spin quadruplet
\begin{equation}
	\begin{aligned}
		\label{eq:di_q}
		&\ket{5,Q_0^\frac{3}{2}} = \op[S\uparrow]{d}^\dag \op[L+\uparrow]{d}^\dag\op[-\uparrow]{d}^\dag \ket{\Omega},\\
		&\ket{5,Q_0^\frac{1}{2}} = \frac{1}{\sqrt{3}} 
  \left( 
  \op[L+\uparrow]{d}^\dag\op[L-\uparrow]{d}^\dag \op[S\downarrow]{d}^\dag 
  -
  \op[S\uparrow]{d}^\dag \op[L-\uparrow]{d}^\dag\op[L+\downarrow]{d}^\dag 
  +
 \op[S\uparrow]{d}^\dag \op[L+\uparrow]{d}^\dag\op[L-\downarrow]{d}^\dag 
 \right) 
 \ket{\Omega},\\
		&\ket{5,Q^{-\frac{1}{2}}} = \frac{1}{\sqrt{3}} 
  \left( 
  \op[S\uparrow]{d}^\dag\op[L+\downarrow]{d}^\dag\op[L-\downarrow]{d}^\dag  
  -
  \op[L+\uparrow]{d}^\dag\op[S\downarrow]{d}^\dag\op[L-\downarrow]{d}^\dag 
  +
  \op[L-\uparrow]{d}^\dag\op[S\downarrow]{d}^\dag\op[L+\downarrow]{d}^\dag 
  \right) 
  \ket{\Omega},\\
		&\ket{5,Q_0^{-\frac{3}{2}}} = \op[S\downarrow]{d}^\dag\op[L+\downarrow]{d}^\dag\op[L-\downarrow]{d}^\dag \ket{\Omega}.
	\end{aligned}
\end{equation}
The energy of this quadruplet is $E_{Q_0} = E_5 -J^\mathrm{ex}_{+-}-2J^\mathrm{ex}_{\rm SL}$. The first excited states form a spin degenerate doublet, with energy  $E_{D_1} = E_5 - J^\mathrm{ex}_{+-} + J^\mathrm{ex}_\mathrm{SL}$. The corresponding eigenstates are given by
\begin{equation}
	\begin{aligned}
		&\ket{5,D_1^\uparrow} = \frac{1}{\sqrt{6}}
  \left(
  2\, \op[L+\uparrow]{d}^\dag \op[L-\uparrow]{d}^\dag \op[S\downarrow]{d}^\dag
  +
  \op[S\uparrow]{d}^\dag \op[L-\uparrow]{d}^\dag \op[L+\downarrow]{d}^\dag
  -
  \op[S\uparrow]{d}^\dag \op[L+\uparrow]{d}^\dag \op[L-\downarrow]{d}^\dag
  \right)
  \ket{\Omega} ,\\
		&\ket{5,D_1^\downarrow}  =  \frac{1}{\sqrt{6}}
  \left(
  2\, \op[S\uparrow]{d}^\dag \op[L+\downarrow]{d}^\dag \op[L-\downarrow]{d}^\dag
  +
  \op[L+\uparrow]{d}^\dag \op[S\uparrow]{d}^\dag \op[L-\downarrow]{d}^\dag
  -
  \op[L-\uparrow]{d}^\dag \op[S\downarrow]{d}^\dag \op[L+\downarrow]{d}^\dag
  \right)
  \ket{\Omega}.
	\end{aligned}	
\end{equation}
The following three energy levels are also spin doublets, with the following eigenenergies 
\begin{equation}
\label{eq:doublets_energies}
E_{D_2} = E_5 -J_{+-}^{\rm p} - J_{\rm SL}^{\rm ex}\,, \quad E_{D_3} = E_5  + J_{+-}^{\rm p} - J_{\rm SL}^{\rm ex}\,, \quad E_{D_4} = E_5  + J_{+-}^{\rm ex} - J_{\rm SL}^{\rm ex}\,.
\end{equation}
The corresponding eigenstates are,
\begin{equation}
	\begin{aligned}
		&\ket{5,D_{2/3}^\uparrow}  = 
  \frac{1}{\sqrt{2}} 
  \left(
  \op[S\uparrow]{d}^\dag\op[L+\uparrow]{d}^\dag\op[L+\downarrow]{d}^\dag 
  \mp
  \op[S\uparrow]{d}^\dag\op[L-\uparrow]{d}^\dag\op[L-\downarrow]{d}^\dag
  \right) \ket{\Omega}, \\
		&\ket{5,D_{2/3}^\downarrow} = \frac{1}{\sqrt{2}} 
  \left(
  \op[L+\uparrow]{d}^\dag\op[S\downarrow]{d}^\dag\op[L+\downarrow]{d}^\dag 
  \mp
  \op[L-\uparrow]{d}^\dag\op[S\downarrow]{d}^\dag\op[L-\downarrow]{d}^\dag 
  \right)\ket{\Omega},
	\end{aligned}
\end{equation}
and, respectively, 
\begin{equation}
	\begin{aligned}
		&\ket{5,D_4^\uparrow} = \frac{1}{\sqrt{2}} 
            \left(
            \op[S\uparrow]{d}^\dag\op[L+\uparrow]{d}^\dag\op[L-\downarrow]{d}^\dag 
            +
            \op[S\uparrow]{d}^\dag \op[L-\uparrow]{d}^\dag\op[L+\downarrow]{d}^\dag 
            \right)
            \ket{\Omega},\\
		&\ket{5,D_4^\downarrow} = \frac{1}{\sqrt{2}} 
            \left(
            \op[L+\uparrow]{d}^\dag \op[S\downarrow]{d}^\dag\op[L-\downarrow]{d}^\dag 
            +
		      \op[L-\uparrow]{d}^\dag \op[S\downarrow]{d}^\dag  \op[L+\downarrow]{d}^\dag 
            \right)
            \ket{\Omega}.
	\end{aligned}
\end{equation}

The first eight states corresponding to the lower energies are not affected by the electron vibron coupling to the modes of symmetry $B_{1g}$. This is evident in \cref{fig:apes_di}, where we observe three parabolas - one is fourfold and the other two are twofold degenerate - with a minimum at $x_1 = 0$. This antisymmetric mode couples instead the eigenstates $\ket{5,D^\sigma_3}$ and $\ket{5,D^\sigma_4}$. For each of the two spin components, the Jahn--Teller Hamiltonian projected on the corresponding two dimensional basis reads 
\begin{equation}
\label{eq:H_JT_dianion}
{\bm H}^{2-}_{\rm JT}(x_1) = \frac{x_1^2}{4} {\bf 1}_2 
- \frac{J_{+-}^{\rm ex} - J_{+-}^{\rm p}}{2}{\boldsymbol \sigma}_z 
+2\frac{g_1}{\sqrt{\hbar \omega_1}}x_1{\boldsymbol \sigma}_x\,. 
\end{equation}
Due to the finite energy splitting in the reference configuration ($x_1 = 0$) this Hamiltonian describes the pseudo-Jahn--Teller effect \cite{Bersuker2006}. The existence of localized minima corresponding to deformed configurations is only guaranteed for a sufficiently large coupling, i.e. $g > \sqrt{\hbar\omega_1 (J_{+-}^{\rm ex} - J_{+-}^{\rm p})}/4$, as can be deduced by analysing the two highest eigenvalues of the dianionic effective Hamiltonian:
\begin{equation}
E_{D_{3/4}}(x_1) = E_5 - J_{\rm SL}^{\rm ex} + \frac{J_{+-}^{\rm ex} + J_{+-}^{\rm p}}{2} + \frac{x_1^2}{4} 
\mp \sqrt{\left(\frac{J_{+-}^{\rm ex}-J_{+-}^{\rm p}}{2}\right)^2 + \frac{4g_1^2}{\hbar\omega_1}x_1^2}\,.
\end{equation}
Differently from the anionic case, here also the eigenstates depend on the deformation coordinate $x_1$. In particular we obtain
\begin{equation}
\label{eq:eigenstates}
\begin{aligned}
\ket{5,D^\sigma_3(x_1)} &= \cos\theta(x_1) \ket{5,D_3^\sigma} -{\rm sgn}\, x_1 \sin \theta(x_1) \ket{5,D_4^\sigma}\,,  \\
\ket{5,D^\sigma_4(x_1)} &= {\rm sgn}\,x_1 \sin \theta(x_1) \ket{5,D_3^\sigma}+\cos\theta(x_1) \ket{5,D_4^\sigma}\,,  \\
\end{aligned}
\end{equation}
where the mixing angle is defined as 
\begin{equation}
\theta(x_1) = \frac{1}{2} \arccos \left(1 + \frac{16 g_1^2 x_1^2}{\hbar \omega_1 (J_{+-}^{\rm ex} - J_{+-}^{\rm p})^2}\right)^{-\frac{1}{2}}\,.
\end{equation}
Within the single mode approximation considered here we obtain a mixing angle at equilibrium $\theta_{\rm eq} \approx 0.24\pi$. The larger is the electron phonon coupling $g$ with respect to the vibrational energy $\hbar \omega_1$ and the energy splitting $J_{+-}^{\rm ex} - J_{+-}^{\rm p}$ of the reference configuration, the larger is the absolute value of the deformation coordinate corresponding to the equilibrium configuration.  Thus, the mixing angle converges to $\theta = \pi/4$. In view of the enhancement of the coupling introduced by the other molecular modes and, above all, by the insulating substrate, we expand the equilibrium eigenstate in the vicinity of this limiting case. 
The lowest energy state having the equilibrium configuration with $x_{1,{\rm eq}} < 0$ reads
\begin{equation}
\begin{split}
\ket{5,D^\sigma_3(x_{1,{\rm eq}})}  &= 
\cos\left(\theta_{\rm eq} - \tfrac{\pi}{4}\right)\frac{1}{\sqrt{2}}\left(\ket{5,D_3^\sigma} + \ket{5,D_4^\sigma}\right) 
- 
\sin \left(\theta_{\rm eq} - \tfrac{\pi}{4}\right)\frac{1}{\sqrt{2}}\left(\ket{5,D_3^\sigma} - \ket{5,D_4^\sigma}\right) \\
&= {\rm sgn}\, \sigma\, \op[S\sigma]{d}^\dag 
\left[
\cos\left(\theta_{\rm eq} - \tfrac{\pi}{4}\right) \op[L_{xz}\uparrow]{d}^\dag \op[L_{xz}\downarrow]{d}^\dag 
+
\sin\left(\theta_{\rm eq} - \tfrac{\pi}{4}\right) \op[L_{yz}\uparrow]{d}^\dag \op[L_{yz}\downarrow]{d}^\dag
\right]
\ket{\Omega}\\
&\equiv \cos \left( \theta_{\rm eq} - \tfrac{\pi}{4} \right) \ket{5,D^\sigma_{xx}} + \sin \left( \theta_{\rm eq} - \tfrac{\pi}{4} \right) \ket{5,D^\sigma_{yy}}\,.
\end{split}
\end{equation}
This state shows a predominant double occupation of the real valued $L_{xz}$ orbital ($\ket{5,D^\sigma_{xx}}$) with a small admixture of the state with a doubly occupied $L_{yz}$ orbital ($\ket{5,D^\sigma_{yy}}$). The roles of $L_{xz}$ and $L_{yz}$ are exchanged if $x_{1,{\rm eq}} > 0$. 

The experimental results we showed in Fig.~1 of the main text suggest that the dianionic ground state has both electrons in the same real valued LUMO. This is clearly not the case in our theoretical model if we consider one mode since the ground state is in the reference configuration where the lowest lying states are the quadruplets which have one electron in each LUMO. We can estimate the reorganization energy required to achieve the desired dianionic ground state by first considering that $E_{D_3} - E_{Q_0} = J_{+-}^{\rm ex} + J_{+-}^{\rm p} + J_{SL}^{\rm ex} = 435$ meV. The total reorganization energy scales with the square of the electron-phonon coupling. Thus, the one associated to the $B_{1g}$ modes scales with the square of the electronic unbalance in the real valued LUMOs. It is thus necessary to achieve a reorganization energy per electron due to $B_{1g}$ modes of least $\SI{109}{\meV}$ for the state $\ket{5,D^\sigma_3(x_{1,{\rm eq}})}$ to become the dianionic ground state.

\section{Combination of multiple modes with the same symmetry}
\label{sec:modes_combination}

In this section, our focus is on combining multiple modes that belong to the same irreducible representation into a single effective mode. This allows us to analyze the potential energy of the system with contributions from all modes of  the same irreducible representation. The combined system of molecule and salt has $C_{4v}$ symmetry. The $B_{1g}$ modes we calculated for the molecule are all in plane modes and thus they transform as $B_1$ modes with respect to the $C_{4v}$ point group. Therefore, we can not only combine the molecular modes among themselves but also include salt contributions to them. We begin this procedure by introducing the transformations of the displacements and momenta given by
\begin{equation}
		\label{eq:trafo_arb}
		\hat{Q}_n = \sum_{\beta \in \{B_1\}}  A_{n\beta} \hat{x}_\beta, \quad
		\hat{P}_m = \sum_{\beta \in \{B_1\}}  B_{m\beta} \hat{p}_\beta. 
\end{equation}
Imposing canonical commutation relations on these new coordinates yields
\begin{equation}
	\label{eq:canonical_trafo}
	\left[ \hat{Q}_n, \hat{P}_{m}\right] = \sum_{\beta\beta'} A_{n\beta}B_{m\beta'} \left[ \hat{x}_\beta,\hat{p}_{\beta'} \right] = 
	i \hbar  \sum_{\beta\beta'} A_{n\beta}B_{m\beta'}\delta_{\beta\beta'} = 
	i\hbar \sum_\alpha A_{n\beta}B_{m\beta} \overset{!}{=} i\hbar\delta_{nm}.
\end{equation}
We further assume $A$ to be orthogonal and thus deduce $B=A$. Moreover, we fix the first row of the transformation matrix to be
\begin{equation}
	A_{1\beta} = \frac{1}{g} \sqrt{\frac{g_\beta^2}{\hbar\omega_\beta}},
\end{equation}
with $g = \sqrt{\sum_\beta \frac{g_\beta^2}{\hbar\omega_\beta}}$. This change of coordinate allows us rewrite the Jahn--Teller Hamiltonian as 
\begin{equation}
\label{eq:h_JT_eff}
	\hat{H}_{JT}(\{Q_n\}) = \frac{1}{4}\sum_n Q_n^2 + g Q_1 \left(\hat{n}_{xz} - \hat{n}_{yz}\right)\,,
\end{equation}
in which the electron-phonon coupling only concerns one effective JT mode, with effective coupling strength $g$.
However, this transformation comes at the cost of a non-diagonal kinetic part in the Hamiltonian. We are still free to construct the matrix transformation ${\bm A}$ such that all the Jahn--Teller inactive modes remain orthogonal among themselves. We can thus consider them as phonon bath for the JT active mode $Q_1$ and we neglect them  in our further discussions. This is justified as far as we are not interested in the dynamics of the JT distortion, but rather on the new equilibrium configurations.

\begin{figure}[ht]
	\centering
	\includegraphics[width=0.8\textwidth]{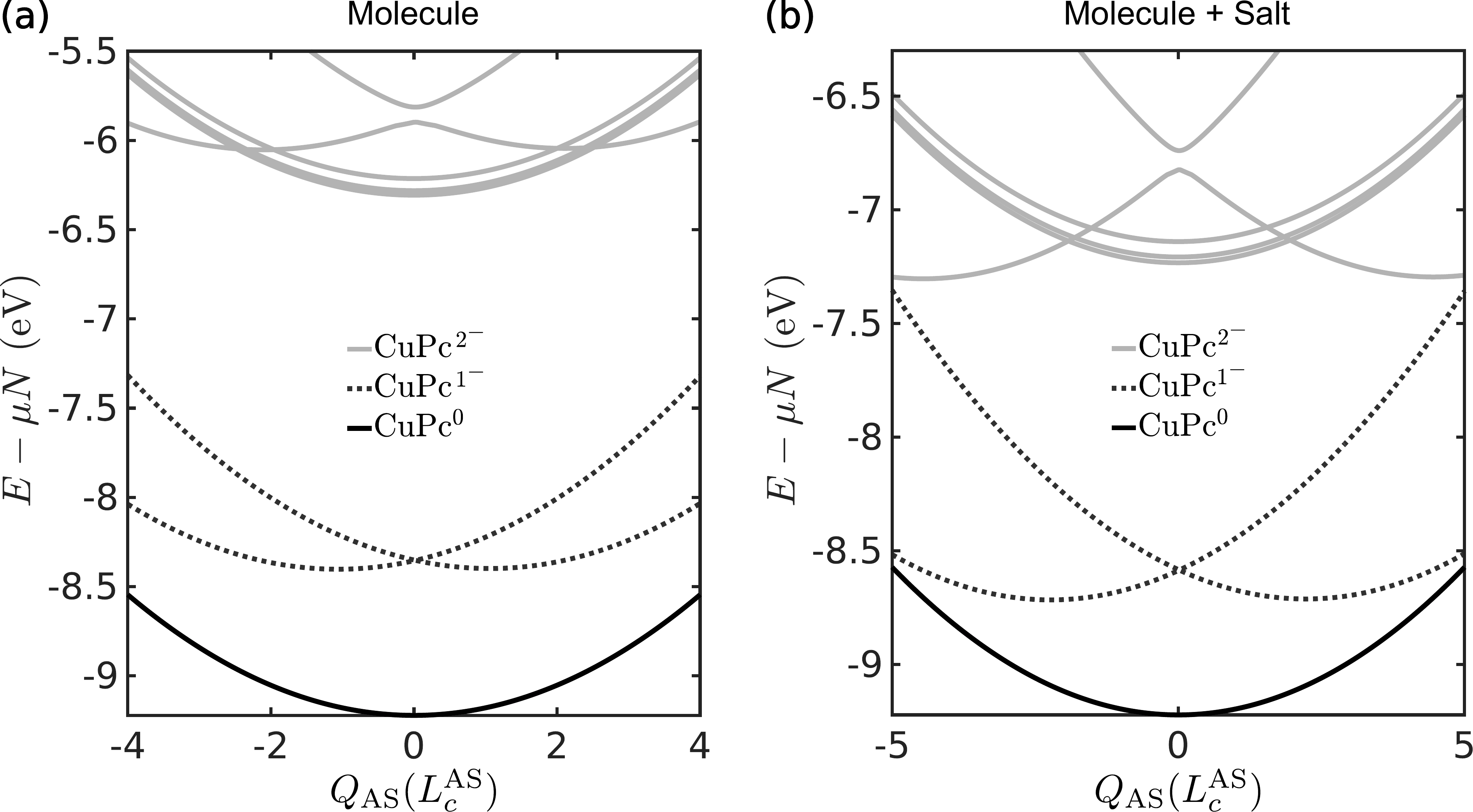}
	\caption{APES of the neutral, anionic and dianionic molecule along the displacement coordinate of the active JT mode. (a) Considering all molecular modes is not enough to obtain the deformed states as ground states.
    (b) By considering all molecular modes combined with a contribution from the salt substrate the states $\ket{5,D^{\sigma}_{xx}}$ become the ground state of the dianionic molecule.} 
	\label{fig:apes_full}
\end{figure}

Only considering molecular contributions to the effective coupling strength yields a value of $g=\SI{218}{\sqrt{\meV}}$ corresponding to a reorganization energy of $\Delta E_{B_1} = \SI{48}{\meV}$. From our previous discussion in \cref{sec:Many-body states} we know that this reorganization energy is not enough for the states with two electrons in one LUMO to become the new ground states. We can also deduce this from the APES in \cref{fig:apes_full} (a), which is plotted along the effective JT coordinate $Q_{\rm AS} := Q_1$ in units of characteristic quantity
\begin{equation}
L^{\rm AS}_c = \sqrt{\frac{1}{g}\sqrt{ \sum_\beta \hbar\omega_\beta g_\beta^2}}\,,
\end{equation}
where $\beta$ runs over all considered modes. We notice that an effective Huang-Rhys factor can be defined, starting from Eq.~\eqref{eq:h_JT_eff}, for the JT active mode $Q_{\rm AS}$  as half of the ratio between the deformed equilibrium configuration $Q_{{\rm AS},\rm min}$ of the anion and the characteristic scaling constant $L^{\rm AS}_c$. In other terms, it reads 
\begin{equation}
 \lambda^{\rm AS}_{\rm eff} = \frac{g}{L^{\rm AS}_c} = \left[ \frac{\left(\sum_\beta\frac{g_\beta^2}{\hbar \omega_\beta}\right)^3}{\sum_{\beta'} \hbar\omega_{\beta'} g_{\beta'}^2}\right]^{1/4}\,.
 \end{equation}
 The latter reduces to the common expression $\lambda = g_\beta/(\hbar\omega_\beta)$ when the sums are restricted to a single mode. 
 The deformed minima in the dianionic configuration are closer to the dianionic minimum energy but this still is not enough to explain the double charging of the same LUMO.
 However, if we consider also the contributions from the salt modes with the parameters given in the main text, we obtain $g = \SI{359}{\sqrt{\meV}}$, yielding a reorganization energy of $\Delta E_{B_1} = \SI{129}{\meV}$. We show the corresponding APES along the $Q_{\rm AS}$ displacement coordinate in \cref{fig:apes_full} (b). For the neutral and anionic charge states we observe no qualitative changes. For the dianionic molecule we see instead a qualitative change, as $\ket{5,D^{\sigma}_{xx}}$ becomes the new global ground state.
The next highest lying states are the $\ket{5,D^{\sigma}_{yy}}$ states. These two pairs of states are split by $2\delta_{\rm as}$, due to the  asymmetry which we included in our Hamiltonian. Otherwise, analogously to the anionic pairs of triplet and singlet states, they would be degenerate.
The energy splitting between the dianionic ground states $\ket{5,D^{\sigma}_{xx}}$ and the quadruplet states is, with our choice of parameters, $\SI{70}{\meV}$. This analysis confirms that the salt substrate in this system has an effect on the charge distribution within a single molecule going beyond mere stabilization of extra charges. We notice, moreover that, even with the contribution of the salt, $\lambda_{\rm eff} \approx 1$ thus allowing for direct transitions between the vibrational ground state with different charges. The classical limit of non-adiabatic vertical transition does not apply to the JT active mode. Analogous considerations can be reserved also to the symmetric ($A_1$) modes, yielding a significant difference in the effective Huang--Rhys factors
 \begin{equation}
\lambda_{\rm eff}^{\rm S} = 2.36\,,\quad \lambda_{\rm eff}^{\rm AS} = 1.11\,.
 \end{equation}

\section{Microscopic model of the salt electron-phonon coupling}%
\label{sec:Microscopic model of the salt electron-phonon coupling}
We present in this section a simple microscopic model to estimate the electron-phonon coupling between an excess charge on the molecule and surface phonons of the underlying salt. We consider the bare Coulomb interaction between the excess electrons in the molecule and the salt ions. Comparison with experimental results indicate that this method seems to overestimate the values for the coupling.
\subsection{Linear vibronic coupling values}
We consider vibration of the ions in all three space directions, thus accounting for  both the longitudinal and the transversal phonon modes in the salt. For simplicity, we limit ourselves to an Einstein model of independent harmonic oscillators with the energies $\hbar\omega_L = \SI{32}{\meV}$ for the vibrations parallel to the surface and $\hbar\omega_T = \SI{20}{\meV}$ for the ones perpendicular to it. The linear electron-phonon  couplings are calculated from the change of the Coulomb interaction between the LUMOs and the salt ions upon displacing of the latter. Specifically we evaluate
\begin{equation}
\label{eq:el-ph_atom}
	\left[{\bm g}_{l}^\alpha\right]_{ij} = \frac{q_l e^2}{4\pi\epsilon_0} 
   \pdv*{\int \: d\bm{r} 
	\frac{\phi^*_{i}(\bm{r})\phi_{j}(\bm{r}) \Delta^\alpha_{0l}}{\left|\bm{r}-\bm{R}^0_l-\bm{u}_l\right|}}
    {u_{l\alpha}}_{\{u_{l\alpha}\}=0}.
\end{equation}
Here $l$ labels an atom in the salt, $i,j = L\pm$ specify one of the degenerate LUMOs and $\alpha = x,y$ or $z$ represents the spatial direction of the displacement. Furthermore, we have $q_i = \pm 1$ depending on the atom being a chloride (+1) or sodium (-1) atom, $e$ is the elementary charge and $\epsilon_0$ the vacuum permittivity.
The atom's equilibrium position is specified by $\bm{R}^0_l$, $\bm{u}_l$ is the displacement coordinate of the considered atom and we have introduced the zero point fluctuations $\Delta^{x/y}_{0l} = \sqrt{\frac{\hbar}{2m_l\omega_L}}$ and $\Delta^z_{0l} = \sqrt{\frac{\hbar}{2m_l\omega_T}}$.
The central copper atom of the molecule, which we define to be at the origin, is centered on a chloride ion and we take 20 layers of salt into account. We checked that the integral in Eq.~\eqref{eq:el-ph_atom} is converged in the $z$ direction (perpendicular to the salt surface) when taking into account this number of layers. In the $x$ and $y$ direction we consider a square with an area of $\SI{400}{\angstrom}^2$, which again suffices for the convergence of the electron phonon couplings. Moreover, we choose a distance of $\SI{2.8}{\angstrom}$ between salt and molecule, which is consistent with values found in the literature \cite{Fatayer2018, HernangomezPerez2020}. By calculating the derivative in Eq.~\eqref{eq:el-ph_atom}, we obtain the expression
\begin{equation}
    \label{eq:salt_coupling}
    \left[{\bm g}_{l}^\alpha\right]_{ij} = \frac{q_l e^2}{4\pi\epsilon_0} \int\:dxdydz
    \frac{\phi^*_{i}(x,y,z)\phi_{j}(x,y,z)(\alpha-R^0_{l\alpha})\Delta^\alpha_{0l}}
    {\left[(x-R^0_{lx}) ^2+(y-R^0_{ly})^2+(z-R^0_{lz})^2\right]^{\frac{3}{2}}}\,,
\end{equation}
which represents the starting point for the numerical calculation of the coupling constants.
\subsection{Numerical evaluation of the integral}
The numerical integration of \cref{eq:salt_coupling} faces the problem of a divergent integrand when the integration variable coincides with one of the equilibrium positions of the atoms in the salt $\bm{r} = \bm{R}_l$. The integral is though not diverging. To better understand this issue, we expand the product of the wave functions around the diverging point as
\begin{equation}
	\begin{aligned}
	\label{eq:psi_expand}
	\phi^*_i(\bm{r})\phi_j(\bm{r}) =  &\phi^*_i(\bm{R}_l^0)\phi_j(\bm{R}_l^0) + 
	\sum_{\alpha}\left(\frac{\partial}{\partial r_\alpha}\phi^*_i(\bm{r})\phi_j(\bm{r})\right)_{{\bm r} = {\bm R}^0_{l}} \left( r_\alpha - R^0_{l\alpha} \right) + \\
		&\frac{1}{2}\sum_{\alpha}\sum_{\beta}\left(\frac{\partial^2}{\partial r_\alpha \partial r_\beta} \phi^*_i(\bm{r})\phi_j(\bm{r})\right)_{{\bm r} = {\bm R}^0_{l}} 
		\left( r_\alpha - R^0_{l\alpha} \right) \left( r_\beta - R^0_{l\beta} \right) + ...,
\end{aligned}
\end{equation}
where $\alpha,\beta = x,y$ or $z$. We now evaluate the integral \cref{eq:salt_coupling} for the first few terms of this expansion in the vicinity of a diverging point ${\bm r} = {\bm R}_l^0 $.
Furthermore, we drop all constants since we are only interested in the behavior of the integral when the integrand is diverging and without loss of generality set $\alpha = z$.
By shifting to cylindrical coordinates with $\zeta = z-R_{l z}$ and $\rho = \sqrt{(x-R_{l x})^2 + (y-R_{l y})^2}$ and integrating over a small volume around the diverging point, we obtain, for the constant term in Eq.~\eqref{eq:psi_expand}
\begin{equation}
	\int_{{0}}^{{2\pi}} {} \: d{\phi} \int_0^R \: d{\rho} \int_{-\epsilon}^\epsilon \: d{\zeta} \frac{\rho \zeta}{\left( \rho^2+\zeta^2 \right)^\frac{3}{2}} = 0.
\end{equation}
For the linear component of \cref{eq:psi_expand}, we distinguish two cases. The first case is for $\alpha = z$, in which it is convenient to choose spherical coordinates with $r = \sqrt{\rho^2 +\zeta^2}$ and $\theta = \arctan \left( \frac{\rho}{\zeta} \right) $. We thus express the integral in the vicinity of the diverging point in these coordinates and obtain 
\begin{equation}
	-\int_0^{2\pi}\: d\phi \int_0^\pi \: d{\theta}\int_0^R \: d{r} \:r \cos^2(\theta)\sin(\theta)	,
\end{equation}
which is not diverging and vanishes in the limit $R \to 0$. The second case we need to discuss is when $\alpha\neq z$. We can choose $\alpha=x$ as the case with $\alpha = y$ is completely analogous.
Now, we rotate our coordinate system and obtain $\zeta' = y-R_{l y}$ and $\rho' = \sqrt{(x-R_{l x})^2 + (z-R_{l z})^2}$.
With this transformation, the local integral is proportional to
\begin{equation}
\int_{{0}}^{{2\pi}} {} \: d{\phi} \int_0^R \: d{\rho'} \int_{-\epsilon}^\epsilon \: d{\zeta'} \frac{\rho'^3 \cos(\phi)\sin(\phi)}{\left( \rho'^2+\zeta'^2 \right)^\frac{3}{2}} = 0 .
\end{equation}
Analogously, one can prove that also the integrals involving higher orders expansions will be finite and vanish in the limit of infinitesimal integration volume around the point with diverging integrand. Therefore, in the numerical evaluation we can set the integrand at the diverging point to 0 and can just use standard numerical integration schemes. 
The electron-phonon coupling relative to each vibrational mode of a salt ion is a Hermitian 2x2 matrix which can be expanded in terms of the Pauli matrices. We obtain 
\begin{equation}
        \bm{g}_l^\alpha = g^\alpha_{l,A_1} {\bm 1}_2 + g^\alpha_{l,B_1}{\bm \sigma}^x+g^\alpha_{l,B_2}{\bm \sigma}^y.
\end{equation}
as, due to the form of the LUMO orbitals, the ${\bm \sigma}^z$ component vanishes exactly. We labelled the different electron-vibron coupling components according to the representations of the corresponding irreducible tensor operator with respect to the $C_{4v}$ point symmetry group. In fact,

\begin{equation}
\sum_\sigma\sum_{ij} \op[Li\sigma]{d}^\dag \delta_{ij} \op[Lj\sigma]{d} \in A_1\,, 
\quad 
\sum_\sigma\sum_{ij}  \op[Li\sigma]{d}^\dag \sigma^x_{ij} \op[Lj\sigma]{d} \in B_1\,, 
\quad \text{and} \quad 
\sum_\sigma\sum_{ij}  \op[Li\sigma]{d}^\dag \sigma^y_{ij} \op[Lj\sigma]{d} \in B_2\,. 
\end{equation}

So far we have shown how to calculate the electron-phonon couplings for each individual atom in the salt. We are, however, more interested in the collective behavior of the substrate. Thus, despite of the simplified Einstein model, we look for the modes with finite coupling to the excess charge in the molecular LUMOS. Analogously to the scheme proposed in Sec.~\ref{sec:modes_combination} we thus combine all modes with the same symmetry. We start with collective transversal modes, whose coordinates read
\begin{equation}
    \begin{aligned}
		x_{\Gamma}^T = \sum_l \frac{g^z_{l,\Gamma}}{\sqrt{\sum_{m} \left(g^z_{m,\Gamma}\right)^2}} \tilde{u}_{l z},\quad
    \end{aligned}
\end{equation}
being $\Gamma = A_1,\,B_1,$ and $B_2$ the corresponding irreducible representation and $\tilde{u}_{l z} = \frac{u_{l z}}{\Delta^z_{0l}}$ the displacement of each individual atom in $z$ direction normalized by the respective zero point fluctuation. Analogously, we define for the longitudinal modes the collective mode coordinates
\begin{equation}
		x_{\Gamma}^L = \sum_l \frac{g^x_{l,\Gamma}\tilde{u}_{l x} + g^y_{l,\Gamma} \tilde{u}_{l y}}
		{\sqrt{\sum_m \left(g_{m,\Gamma}^x\right)^2 + \left(g_{m,\Gamma}^y\right)^2}}\,.
  \end{equation}
These coordinate transformations identify six collective salt modes with a finite electron phonon coupling. All the other modes are not influenced by an excess charge on the CuPc molecule and are treated as a dissipative phonon bath.
The coupled modes exhibit respectively the transversal and longitudinal coupling strengths $g^T_\Gamma = \sqrt{\sum_l \left(g^z_{l,\Gamma}\right)^2}$ and $g^L_\Gamma = \sqrt{\sum_l \left(g^x_{l,\Gamma} \right)^2+\left(g^y_{l,\Gamma}\right)^2}$, with $\Gamma = A_1, B_1$ and $B_2$. Starting from the numerical calculation of the integrals in \eqref{eq:salt_coupling}, one obtains the following coupling strengths for the collective modes
\begin{equation}
	\begin{aligned}
		g^T_{A_1} &= \SI{116}{\meV}, \quad &&g^T_{B_1} = \SI{20.6}{\meV}, \quad &&g^T_{B_2} = \SI{2.5}{\meV}\\
		g^L_{A_1} &= \SI{92.7}{\meV}, \quad &&g^L_{B_1} = \SI{11.1}{\meV}, \quad &&g^L_{B_2} = \SI{11.7}{\meV}.
	\end{aligned}
\end{equation}
Similarly to what we observed with the molecular modes, also in the salt the contribution to the electron-phonon coupling of the $B_1$ mode dominates over the one of the $B_2$ modes. Therefore, we concentrate our discussion solely on the $A_1$ and $B_1$ modes. They yield a total reorganization energy
\begin{equation}
	\Delta  E_\mathrm{reorg}^\mathrm{salt} = \frac{\left(g^T_{A_1}\right)^2+\left(g^T_{B_1}\right)^2}{\hbar\omega_T}+
	\frac{\left(g^L_{A_1}\right)^2+\left(g^L_{B_1}\right)^2}{\hbar\omega_L} = \SI{966}{\meV}.
\end{equation}
These values of the reorganization energy due to the salt deformation can be taken as an estimate for their order of magnitude. Their quantitative accuracy is though problematic.

On the one hand, by following the conclusions from the experiments of Fatayer et al. \cite{Fatayer2018} we expect that the total reorganization energy upon charging should be in the range of $\SI{300}{\meV} \leq \Delta E_\mathrm{reorg}^\mathrm{tot} \leq \SI{500}{\meV}$. As the molecular component is $\Delta E_\mathrm{reorg}^\mathrm{mol} = \SI{187}{\meV}$, a salt reorganization energy in the range $\SI{113}{\meV} \leq \Delta E_\mathrm{reorg}^\mathrm{salt} \leq \SI{313}{\meV}$ is to be expected. 
This leads us to the conclusion that our model overestimates the total reorganization energy stemming from the salt.

On the other hand, in the molecular reference condition the energy difference between the dianionic ground state (both electrons in different LUMOs) and the localized configuration is $\approx \SI{200}{\meV}$.
Thus the minimum reorganization energy stemming from the antisymmetric modes, in order to obtain the localized configuration as the dianionic state with lowest energy, must be $\approx \SI{100}{\meV}$.

We assume that these problems can be solved by a more realistic treatment of the salt and salt-molecule interaction which includes the dispersion relation of the phonons and a screening of the bare Coulomb interaction between the molecule and the underlying ions. While keeping the form of the Hamiltonian Eq.~\eqref{eq:h_eff_1mode}, we have considered the effective couplings as free parameters and explored their effects on the many body transitions. 
We have chosen them to be $\Delta E_{A_1}^\mathrm{salt} = \SI{231.2}{\meV} $ and $\Delta E_{B_1}^\mathrm{salt} = \SI{80}{\meV} $.
Combined with the molecular contributions this choice of parameters gives a total reorganization energy of $\Delta E_{A_1} + \Delta E_{B_1} = \SI{500}{\meV}$, thus not overestimating the total reorganization energy while having the localized dianionic configuration as the ground state.


\end{document}